\documentclass[jeosman,preprint,fleqn,showpacs,showkeys]{revtex4}
\usepackage{graphicx}
\usepackage{amssymb,amsmath,bm}

\begin{document}

\title{Can a skier make a circular turn without any active movement?}

\author{ Sun Hyun Youn  \footnote{E-mail: sunyoun@jnu.ac.kr, fax: +82-62-530-3369}}
\address{Department of Physics, Chonnam National University, Gwangju 500-757, Korea}

\begin{abstract}

  A skier's motion was analyzed by a simple model consist of point mass $m$ and
  a single rod connected to a single ski plate. We studied the conditions for the stable ski turn as
functions of the linear velocity and the radius of the turn. The
solutions for the stable ski turn in our model do not require any
extra skier's movement to complete a stable circular turn. The
solution may then give the skier the most comfortable skiing method
without any active movement to control the ski. The generalized
force supporting the point mass from the ski plate was calculated.
We obtained the force from the ground (rebound force) without any
geometrical structure of the ski plate. Adding an active movement to
the direction  of the ski plate,  the conditions for the stable ski
turn were also analyzed.  Our result gives some insight for the
skier who wants to develop technique.



\end{abstract}


\maketitle

\section{Introduction}

   The skill of the alpine ski has developed over a long time.
Classical alpine ski styles, such as the American style, Austrian
style, and French style, have been briefly introduced from the
physicist's point of view\cite{shonle1972}. They explained the basic
dynamics of rotations in alpine ski from torque equations. Without
external torque, a ski can make turns  with counter-rotations by
unweighting. In the Austrian style, a vigorous unweighting is
applied to initiate the turns. On the other hand, the American style
uses less counter-rotation and more external torque by edging
control. The French style makes  strong use of the external torques
produced by having the skier's weight backwards on an edged ski to
commence a turn.

 The dynamics of ski skating has also been studied with mathematical
model of ski skating on a level plane.\cite{driessel2004}. They
obtained the maximized averaged speed for a given power. They showed
that the skating technique, where the ski moves at an angle to the
direction of motion, is much faster than the classical technique,
 in which the ski stays in a track parallel to the direction of motion.

As the measurement technology developed, analysis of ski technique
was carried out. Digitizing synchronized video sequences, the ski
turn techniques of experienced and intermediate skiers were analyzed
with measuring the selected kinematic variables for
 different ski turns\cite{muller1998}.  In 2000, the carving ski
 turn was studied with specially developed joint angle sensors
 attached to the skier's leg, and force sensors fitted between the
 binding plate and the ski\cite{yoneyama2000}. The joint motion
 using the carving ski is moderate, with no impact resistance. In contrast,
 in a short turn using a conventional ski, the skier makes a quick
 rotation of the thigh in the early part of the turn. As a result, the reacting
 force occurs instantly, and its amplitude of change is larger.

Using analytical techniques including electromyography, kinetic, and
kinematic methods and computer simulations, new insights into the
skills of both alpine skiing and ski-jumping have been studied
\cite{muller2003}.  Biomechanical  aspects of new techniques are
analyzed according to the specific conditions in alpine skiing and
the effects of equipment and individual specific abilities on
performance. In 2012, kinetic analysis of ski turns in alpine skiing
based on the measured ground reaction forces were
published\cite{veverka2012}.  Their method was based on a
theoretical analysis of physical forces acting during the ski turn.
They defined two elementary standard phases of a ski turn by
analyzing the external forces acting on the system during a turn.
The initiation phase is for the preparing to turn and the steering
phase is for the actual turning, during which the center of gravity
of the skier-ski system moves along a curvilinear trajectory.

  Other recent studies on skiing include a machine learning algorithm to
detect and label turns in alpine skiing using a single sensor placed
on a skier's knee \cite{jones2016}. The development of an
interactive ski-simulation motion recognition system by
physics-based analysis has also been  studied\cite{jin2013}.

The development of skiing technique  from the ancient era has been
studied until now. Some works explained how the ski makes turns, and
what is the optimum path theoretically. Other works analyzed the
skier's motion by measuring the position, and forces applied to the
skier. In this article we try to analyze  the skier's motion with a
point mass $m$ on the
   single rod connected to a single ski plate. This model excludes the
   skier's active motion,  and finds the particular solution for the point mass $m$
   to make stable ski turns. This solution may give the skier the
   most comfortable skiing method without any active movement to
  control the ski. We also assumed that the ski plate moves following a circular trajectory on
    an inclined slope.

  The present paper is organized as follows: Section \ref{motion}
  sets the Lagrangian for a point mass $m$ on a
massless rod (length $l_0$) that is connected to a point mass and a
single ski plate. The ski plate moves following a circular
trajectory on the inclined slope. We assumed that the mass of the
ski plate is zero, and we neglect any internal movement of the ski
plate. We include two constraints in order to find two generalized
forces. The first one keeps the ski plate moving along a circular
path on the slope, and the second one keeps the rod attached to the
ski plate. For the skier's point of view, the second generalized
force is related to the rebound force, or the force from the ground
to the skier.

Section \ref{secExample} solves the differential equations
numerically with setting the angle $\beta$ as a constant value, and
finds the time-dependent function of the angles $\alpha$ and
$\theta$.  The initial condition of two angles $\alpha$ and $\beta$
is studied for both making a complete circular turn without falling,
and making a successive turn. In general, there is no initial
condition that the final condition at $\theta = 90^{\circ}$ is
another new initial condition for the next turn. Since the skier can
adjust her/his motion by her/his two legs and poles,  we checked the
angular velocity of $\theta$ at first, in order to make the same
initial condition for the next turn in this section. We found
solutions for the functions of the angle of the slope, the speed of
the ski,  and the radius of the turn.

On the slope, a skier usually actively moves her/his body in order
to keep a stable position and control the ski path. Section
 \ref{secAvtive} includes the skier's active motion
 parallel to the ski plate in our model, while in Section
\ref{secExample}, our model moves passively during the ski turn, in
other words, once the initial angles $\alpha_i$ and $\beta_i$ are
chosen, the point mass $m$ on the bar  moves down hill with the same
angle $\beta_i$ along the circular ski trajectory, without any other
active movement. In Section \ref{secAvtive}, the angle $\beta$ is
not constant during a ski turn. We set the angle $\beta$ as a
time-dependent function, which can duplicate the skier's motion
along the ski plate back and forth. We analyzed the motion of the
point mass $m$ that corresponds to the center of the skier's mass as
the angle $\beta$ changes as a function of time.

Section \ref{conclusion} summarize the  main results, and discusses
their application.

\section{Lagrangian for the skier's simplified motion.} \label{motion}

\begin{figure}[htbp]
\centering
\includegraphics[width=10cm]{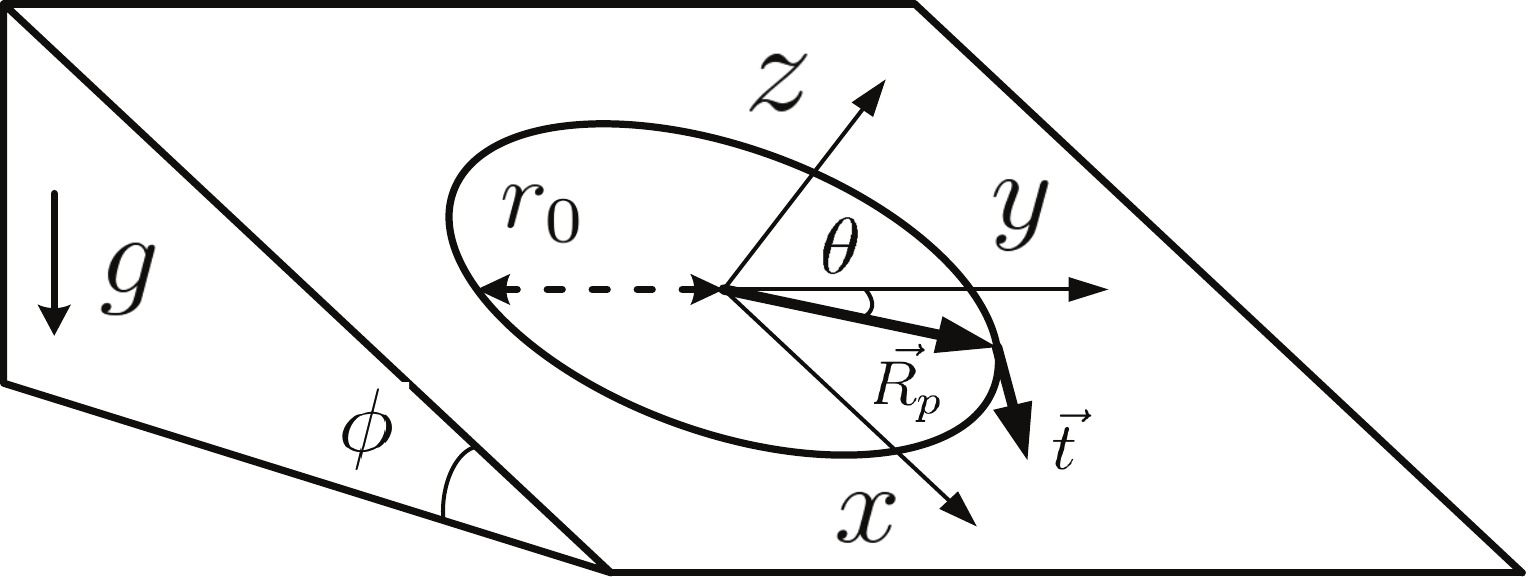}
\caption{Ski trajectory is a circle with a radius $r_0$, and its
center is at the origin of the $xyz$ coordinate system. The $x$-axis
is along the fall line, and the $z$-axis is normal to the slope. The
direction of $g$ is along the gravitational field. $ \vec{R}_p $
represents the position of the ski center and $ \vec{t}$ is along
the ski direction. $\phi$ is the angle between the slope and the
horizontal.} \label{FigSkiXYZ}
\end{figure}

Although a skier moves on two plates with active motions, we studied
the simplified skier's motion as a point mass $m$ on the massless
rod (length $l_0$), which connected the point mass and the single
ski plate. We did not include the motion of the plate, in other
words, we assumed the mass of the ski plate is zero and the ski
plate moves following a circular trajectory on an inclined slope.
 Figure \ref{FigSkiXYZ} defines the coordinates: the origin
of the coordinate is at the center of the circle, and the $x$-axis
is parallel to the fall line, and the $z$-axis is normal to the
slope. The angle $\theta$ is measured from the $y$-axis and the
radius of the circle on the slope is $r_0$. The position of the rod
which is connected to the ski plate moving along a circular
trajectory can be written as follows:
\begin{eqnarray}
\vec {R}_p  &=& \{r(t)  \sin \theta(t), r(t) \cos \theta(t) ,0 \}
\\
\hat{u}&=& \frac{\vec{R}_p } {|\vec{R}_p|},\\
\hat{t}&=&  \hat{u} \times \vec{z}, \label{EqDefinRUT}
\end{eqnarray}
where, $ \hat u$ is the unit vector, and the unit vector $\vec{t}$
is the tangential vector pointing in the direction of motion of the
ski plate.

\begin{figure}[htbp]
\centering
\includegraphics[width=5cm]{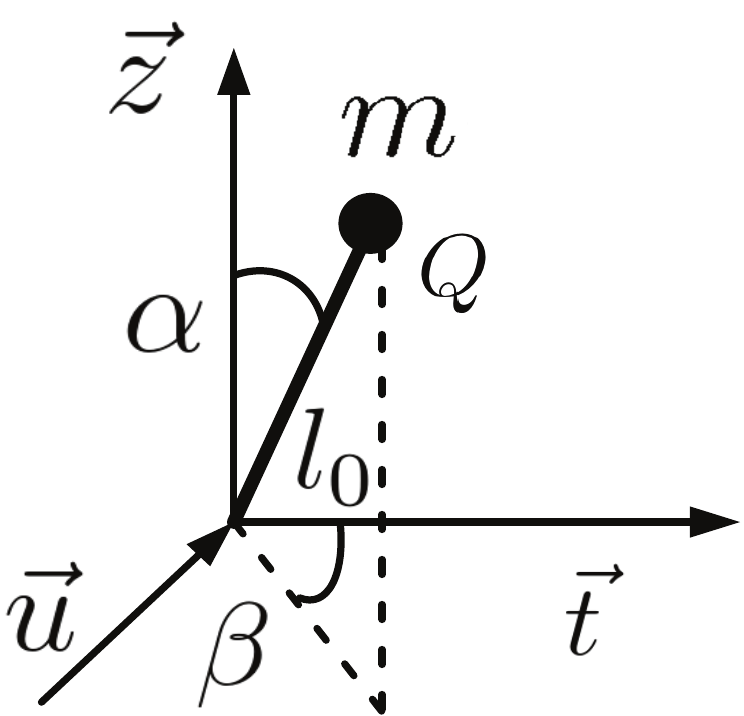}
\caption{The human body is simplified as a massless rod and point
body. The length of the rod is $l_0$ and the mass of the body is
$m$. The unit vector $\vec u$ is  parallel to the radius vector and
the unit vector $\vec t$ is along the ski direction. $\alpha$(
$\beta$) is the angle between the rod and $\vec z$ ($\vec u$).}
\label{FigSkiBody}
\end{figure}

 In Fig. \ref{FigSkiBody}, the
 angle  $\alpha$ is the angle between the $z$-axis and the rod, and
 the angle $\beta$ is the angle between the unit vector $-\vec u$ and
 the projection of the rod on the slope. If angle $\beta$ is zero,
 the angle $\alpha$ represents how much
  the skier body leans toward the center of the circular trajectory.
For a given angle $\alpha$, if the angle $\beta$ increase, the
skier's body leans toward the direction of the ski movement.

The position ($\vec Q $) of the point mass $m$  is;
\begin{eqnarray}
\vec{Q} &=& \vec {R}_p - l_0  \sin \alpha(t) \cos\beta(t) \hat{u} +
l_0 \sin \alpha(t) \sin\beta(t) \hat{t} + (l_0
\cos(\alpha(t)) + b_z(t) )\vec{z}, \\
 h(t) &=& \vec{Q}. \vec { z}  \cos \phi - \vec{Q}.\vec{z}  \sin \phi,   \label{EqDefineQH}
\end{eqnarray}

where, $h(t)$ is the height of the point mass $m$ relative to the
origin of the circular trajectory.

In our model, we can calculate the Lagrangian for the point mass $m$
that moves on the top of the rod.
\begin{eqnarray}
L =  T - V  = \frac{1}{2}  \frac{d {\vec Q}}{ dt} \cdot \frac{d
{\vec Q} }{ dt} - m g h(t). \label{EqLag1}
\end{eqnarray}

In order to calculate the generalized force we used two constraints,
as follow:
\begin{eqnarray}
f_r &=& r(t) - r_0, \\
f_z &=& b_z (t), \label{EqConst1}
\end{eqnarray}

where, $f_r$ can be used to find the generalized force that keeps
the ski plate moving along the circular path on the slope. The other
constraint $f_z$ is also used to check the generalized force that
keeps the rod attached to the ski plate.

The Lagrangian equations give the following five equations of
motion:
\begin{eqnarray}
\frac{d}{dt} ( \frac{\partial L} { \partial \dot{\theta }}) -
\frac{\partial L} { \partial \theta } - \lambda_r  \frac{\partial
f_r} {\partial \theta } -  \lambda_z  \frac{\partial f_z} {\partial
\theta } = 0, \label{EqLagrangeTA}
\end{eqnarray}
\begin{eqnarray}
\frac{d}{dt} ( \frac{\partial L} { \partial \dot{\alpha }}) -
\frac{\partial L} { \partial \alpha } - \lambda_r  \frac{\partial
f_r} {\partial \alpha } -  \lambda_z  \frac{\partial f_z} {\partial
\alpha } = 0, \label{EqLagrangeTB}
\end{eqnarray}
\begin{eqnarray}
\frac{d}{dt} ( \frac{\partial L} { \partial \dot{\beta }}) -
\frac{\partial L} { \partial \beta } - \lambda_r  \frac{\partial
f_r} {\partial \beta } -  \lambda_z  \frac{\partial f_z} {\partial
\beta } = 0,  \label{EqLagrangeTC}
\end{eqnarray}
\begin{eqnarray}
\frac{d}{dt} ( \frac{\partial L} { \partial \dot{r }}) -
\frac{\partial L} { \partial r } - \lambda_r  \frac{\partial f_r}
{\partial r } -  \lambda_z  \frac{\partial f_z} {\partial r } =
0,
 \label{EqLagrangeConA}
\end{eqnarray}
\begin{eqnarray}
\frac{d}{dt} ( \frac{\partial L} { \partial \dot{b_z }}) -
\frac{\partial L} { \partial b_z } - \lambda_r  \frac{\partial f_r}
{\partial b_z } -  \lambda_z  \frac{\partial f_z} {\partial b_z } =
0.
 \label{EqLagrangeConB}
\end{eqnarray}

From  Eqs. \ref{EqLagrangeConA}-  \ref{EqLagrangeConB} , we obtain
$\lambda_r $ and $\lambda_z$ as the following, under the constraint
condition $f_r = 0$ and $f_z = 0 $.
\begin{eqnarray}
\lambda_r  =& &l_0 \cos \beta \sin \alpha \dot \alpha ^2 - m g \sin
\phi \sin \theta +   l_0 \cos \beta \sin \alpha \dot \beta ^2 + 2
l_0 \cos \alpha \sin \beta \dot \alpha ( \dot \beta - \dot \theta )
\nonumber \\
 &-& 2 l_0 \cos \beta \sin \alpha \dot \beta \dot \theta + r_0 \dot
\theta ^2 + l_0 \cos \beta \sin \alpha \dot \theta ^2- l_0 \cos
\alpha \cos \beta \ddot \alpha  \nonumber \\ &+& l_0 \sin \alpha
\sin \beta \ddot \beta - l_0 \sin \alpha \sin \beta \ddot \theta
 \label{EqLambdaR}
\end{eqnarray}
\begin{eqnarray}
\lambda_z  = m g \cos \phi - l_0 \cos \alpha \dot \alpha ^2  - l_0
\sin \alpha \ddot \alpha
 \label{EqLambdaZ}
\end{eqnarray}
Then the generalized force $Q_r$ and $Q_z$  corresponding to the
constraint conditions becomes:
\begin{eqnarray}
Q_r &=&  \lambda_r  \frac{\partial f_r} {\partial r }  +  \lambda_z
\frac{\partial f_z} {\partial r }, \nonumber \\
    &=& \lambda_r,
\label{EqGenFr}
\end{eqnarray}
\begin{eqnarray}
Q_z   &=&   \lambda_r \frac{\partial f_r} {\partial b_z } +
\lambda_z \frac{\partial f_z} {\partial b_z }, \nonumber \\
&=& \lambda_z
 \label{EqLambdaZ}
\end{eqnarray}

Under the constraint conditions, the equations of the motion for the
three variables ( $\theta, \alpha, \beta$) are as follow:

\begin{eqnarray}
& & m g r_0 \cos\theta \sin\phi -
 m g l_0 \cos\beta \cos\theta \sin\phi \sin\alpha -
 m g l_0 \sin\phi \sin\alpha \sin\beta \sin\theta  \nonumber \\
 & + & l_0 r_0 \sin\alpha \sin\beta \dot \alpha ^2 -
 2 l_0 r_0 \cos\alpha \cos\beta \dot \alpha \dot \beta +
 2 l_0 ^2 \cos\alpha \sin\alpha \dot \alpha \dot \beta +
 l_0 r_0 \sin\alpha \sin\beta \dot \beta ^2 \nonumber \\  &+&
 2 l_ 0 r_0 \cos\alpha \cos\beta \dot \alpha\dot \theta
- 2 l_0 ^2 \cos\alpha \sin\alpha \dot \alpha\dot \theta -
 2 l_0 r_0 \sin\alpha \sin\beta \dot \beta\dot \theta -
 l_0 r_0 \cos\alpha \sin\beta \ddot \alpha \nonumber \\ &-&
 l_0 r_0 \cos\beta \sin\alpha \ddot \beta+
 l_0 ^2 \sin\alpha^2\ddot \beta - r_0 ^2 \ddot \theta +
 2 l_0 r_0 \cos\beta \sin\alpha \ddot \theta -
 l_0 ^2 \sin\alpha^2\ddot \theta =0
 \label{EqLambdaTheta}
\end{eqnarray}
\begin{eqnarray}
& & l_0 (m g \cos\phi\sin\alpha +
   m g \cos\alpha \cos\theta \sin\phi\sin\beta -
   m g \cos\alpha \cos\beta \sin\phi\sin\theta -
   l_0 \sin2 \alpha \dot \beta  \dot
    \theta  \nonumber \\ &+&
   l_0 \cos\alpha \sin\alpha \dot \beta ^2  + \cos\alpha (-r_0 \cos\beta + l_0 \sin\alpha) \dot
     \theta ^2 - l_0 \ddot \alpha -
   r_0 \cos\alpha \sin\beta \ddot \theta)=0
\label{EqLambdaAlpha}
\end{eqnarray}
\begin{eqnarray}
& & l_ 0 \sin\alpha (m g \cos\beta \cos\theta \sin \phi +
   m g \sin\phi\sin\beta \sin\theta -
   2 l_ 0 \cos\alpha \dot \alpha (\dot \beta - \dot \theta) \nonumber \\
   &+&
   r_ 0 \sin\beta \dot \theta ^2 - l_ 0 \sin\alpha \ddot \beta -
   r_ 0 \cos\beta \ddot \theta + l_ 0 \sin\alpha \ddot \theta)=0
\label{EqLambdaBeta}
\end{eqnarray}

\section{Skier's initial setup for making stable and successive turns. } \label{secExample}

We found the numerical solutions for the simplified skier's motion.
In our model, the path of the ski is restricted to  a circular orbit
on an inclined slope, and we assumed the skier as a point mass $m$
on the top of a rod with length $\l_0$. We also assumed that the ski
plate is a point at the bottom of the rod. Solving the differential
equations numerically, we set the angle $\beta$ as a constant value
in this section, and found the time-dependent function of the angle
$\alpha$, and $\theta$.

\begin{figure}[htbp]

\centering
\includegraphics[width=10cm]{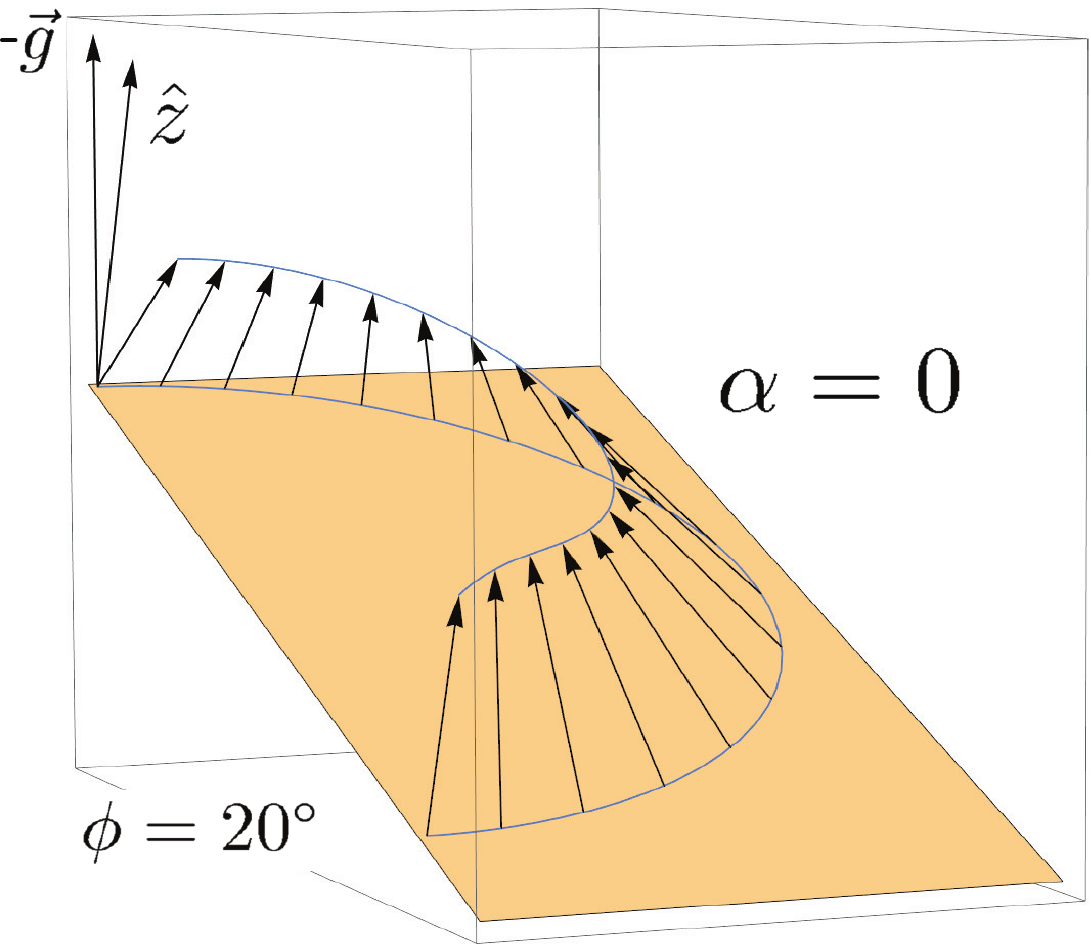}
\caption{The arrow indicates the point $Q$ and the starting points
of arrows follows the circular orbit. The ski moves from $\theta = -
90^{\circ} $ to the angle $\theta =90^{\circ}$. At $\theta
=90^{\circ}$, the angle between the $\hat z$ and the arrow $\alpha $
becomes zero. ($\alpha_i =33.9^{\circ}, \beta_i =8.3^{\circ}$) }
\label{FigSlopeAZero}
\end{figure}

Figures  \ref{FigSlopeAZero} shows the time traces of the point mass
$m$ on $Q$ that moves on the circular trajectory on the slope
inclined by $20^{\circ}$. At $\theta=-90^{\circ}$, the initial
angles $\alpha_i$ and $\beta_i$ are ( $33.9^{\circ}$, $8.3^{\circ}$)
respectively. The initial angular velocity of $\dot{\theta_0} =
\frac{120 \pi}{180}$. We assumed the radius of the circle($ r_0 $)
is $2 {\rm m} $ and the length ($ l_0 $) between the point mass $m$
and the ski is $ 1 {\rm m} $. Although these values are not fitted
to the real situation, this data set is enough to analyze the ski
movement. We also set the angle $\beta(t)$ as a fixed value as
$\beta_i $; if not the motion becomes too complicated and, it is not
easy to find stable ski motion on a circular trajectory.

\begin{figure}[htbp]
\centering
\includegraphics[width=10cm]{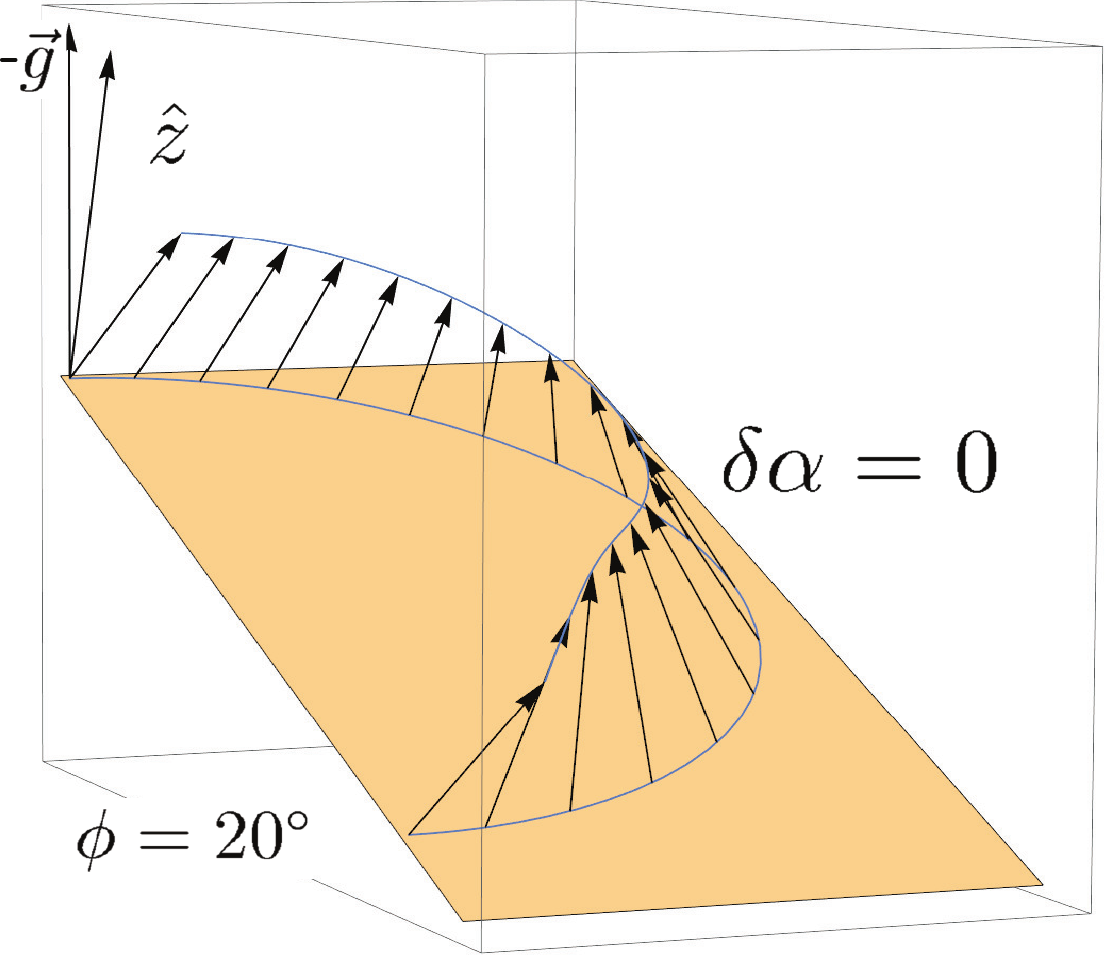}
\caption{The arrow indicates the point $Q$ and the starting points
of arrows follows the circular orbit. The ski moves from $\theta = -
90^{\circ}$ to the angle $\theta =90^{\circ}$. At $\theta
=90^{\circ}$, the angle between $\hat z$ and the arrow $\alpha $ is
the same as the $ -\alpha_0 $ at $t=0$.  ($\alpha_i =29.9^{\circ},
\beta_i =26.0^{\circ}$)   } \label{FigSlopeDAZero}
\end{figure}

 Figure \ref{FigSlopeDAZero} shows new results for the same slope with the
same initial conditions, except for the initial angles $\alpha_i ,
\beta_i $. The initial angles are ($\alpha_i =29.9^{\circ}, \beta_i
=26.0^{\circ}$). The initial conditions in Fig. \ref{FigSlopeAZero}
give the final angle $\alpha_f $ at $\theta = 90^{\circ}$ being
zero, but the initial conditions in Fig. \ref{FigSlopeDAZero} give
$\alpha_f = -29.9^{\circ}$. For the next turn, the center of the
turn is on the left side, so $\alpha = -29.9^{\circ}$ is the same
initial conditions for the angle $\alpha$.


\begin{figure}[htbp]
\centering
\includegraphics[width=10cm]{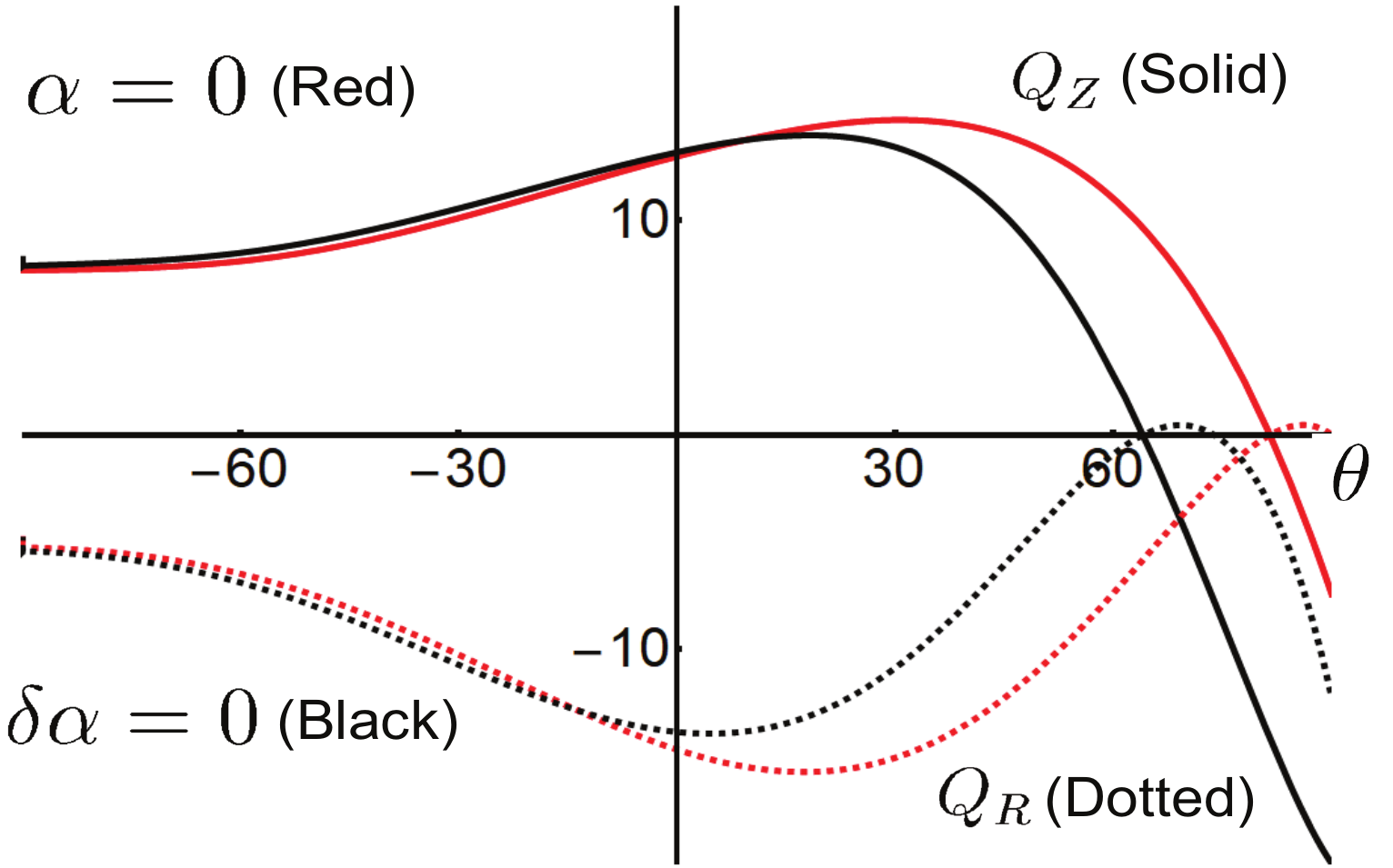}
\caption{The solid (dashed) red line represents the generalized
force $Q_z$ ($Q_R $) for the case where the angle $\alpha$ becomes
zero at $\theta = 90^{\circ}$. The solid (dashed) black line
represents the generalized force $Q_z$ ($Q_R $) for the case where
the angle $\alpha$ becomes $-\alpha_0$  at $\theta = 90^{\circ}$.
 } \label{FigZRforce}
\end{figure}

  In Fig. \ref{FigZRforce}, we plot the generalized force $Q_z $ and
$Q_R $. When the final angle $\alpha = 0 $, the force $Q_z$ remains
positive till $\theta_c = 80^{\circ}$; after that the generalized
force becomes negative.  In other words, there should be a negative
force to keep the circular trajectory motion, which is to say  the
force should be toward the bottom. From the skier's view-point,
he/she feels an uprising force from the ground. In actual skier's
movement, he/she can absorb this force by bending his/her body, or
he/she is going to fall down. We obtained this uprising force
without any motion of the ski plate, in other words, there exists a
force from the ground when the skier moves to the down hill
following a circular trajectory.  For the case to get a final
condition $\delta \alpha =0 $, the generalized force becomes
negative just after $\theta_c = 65^{\circ}$, and the force is
greater than for the case $\alpha=0$ in Fig. \ref{FigZRforce}.
Considering the generalized force $Q_R$, the force is negative till
the angle $\theta = \theta_c$, but the force becomes positive after
the angle $\theta_c$. From the skier's view-point, to follow the
circular trajectory, he/she should generate a force toward the
center of the circle by edging of the ski. However, after the angle
$\theta_c$, the direction of the force is changed. The skier needs a
force outward the center of circle to accomplish the circular
motion. In actual skiing, it's related to the change of the edge
during the ski turn.

\begin{figure}[htbp]
\centering
\includegraphics[width=10cm]{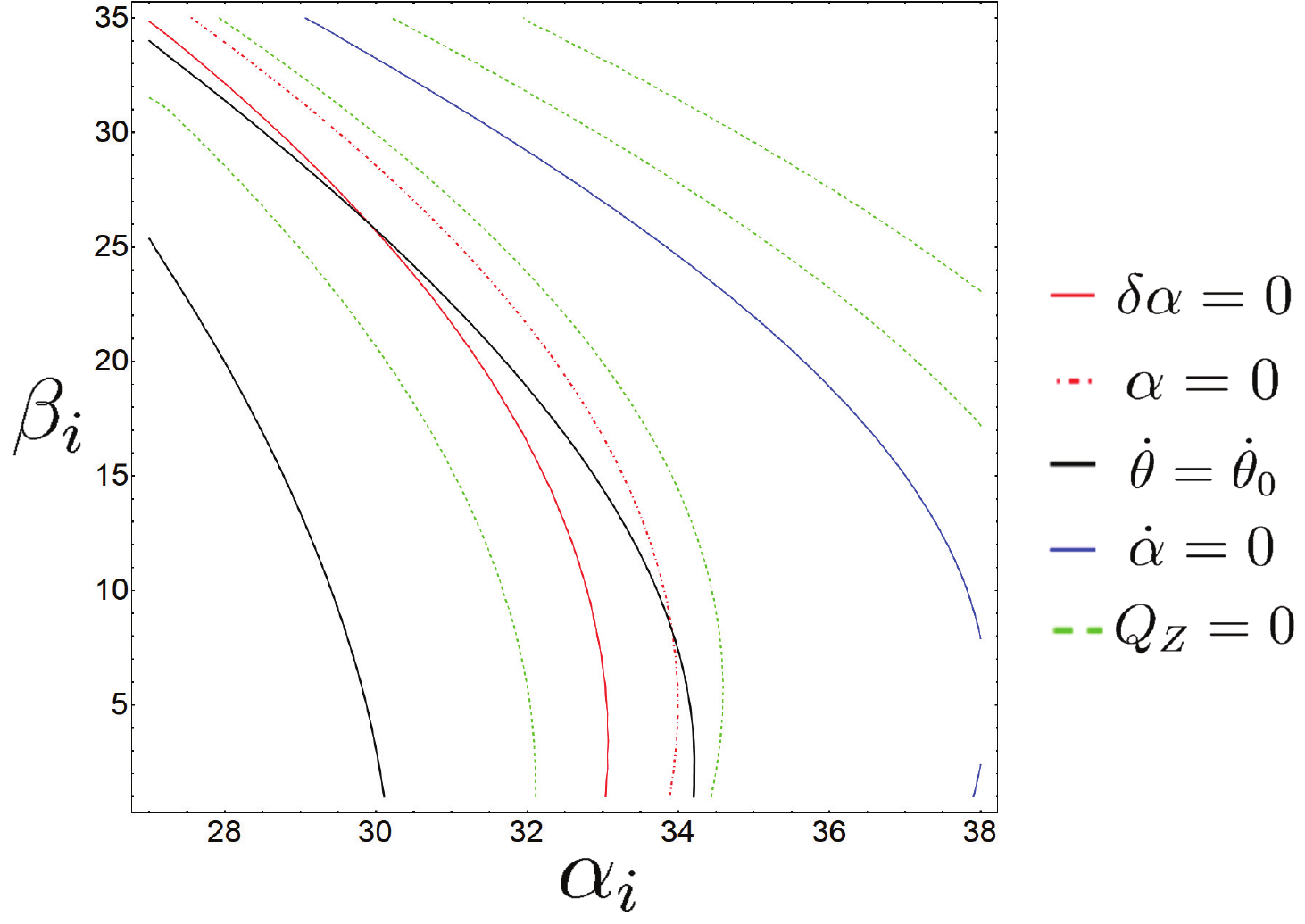}
\caption{Contour lines as a function of the initial angles $\alpha$
and $\beta$.  The solid red line represents the case when the
magnitude of the angle $\alpha$ at $\theta =90^{\circ}$ is the same
as the initial angle. The dashed red line represents the case when
the magnitude of the angle $\alpha$ at $\theta = 90^{\circ}$ is
zero. The solid black line represents the case when the derivative
value of the $\theta$ at  $\theta = 90^{\circ}$ is  the same at the
initial value $\dot \theta_0 $. The solid blue line represents the
case when the derivative value of the $\alpha$ at
 $\theta =90^{\circ}$ is  zero.
 The dashed green line represents the case
when the magnitude of the generalized force $Q_z $ equals zero at
$\theta =90^{\circ}$.   } \label{FigContourA}
\end{figure}

In order to  make
  continuous ski turns without adding extra force, all variables
  at $\theta = 90^{\circ}$ recover the values at $\theta =
  -90^{\circ}$, except the $\alpha_f$. Since the center of the circle
  moves from the right to the left, the $\alpha_f$ should be
  $-\alpha_i$. In Fig. \ref{FigContourA},
  the solid red line represents the initial angles $\alpha_i $ and
  $\beta_i$, which gives the final angles $\beta_f = \beta_i $
  and $\alpha_f = - \alpha_i $. The angles on the contour line  $\delta \alpha =
  0 $ gives the original initial condition for the angle-$\alpha$  at $\theta =
  -90 ^{\circ}$. As the initial angle $\beta_i$ changes from 0 to
$35^{\circ}$, the $\alpha_i$ varies from $33^{\circ}$ to
$27^{\circ}$, in order to give the final angle $\alpha_f
=-\alpha_i$.

 Since we fixed the angle $\beta$ during the turn, we check the
 two variables $\alpha, \theta$ to recover
  the initial conditions. Considering the angle  $\theta$, we already
  fix the time at which $\theta(\tau) = -90^{\circ}$, $\tau$ can be
  an  initial time, and the angle may be set as $\theta= - 90^{\circ}$ at
  the next turn. Therefore the initial angles ($\alpha_i $ , $\beta_i$) related
   to the contour $\delta \alpha = 0$ give
  almost all initial values for the second turn.

  On the other hand,
  if we check the angular velocity of two variables $\alpha$ and
  $\theta$, the two angular velocities are different from the initial
  angular velocities. The solid black line in Fig.
  \ref{FigContourA}, represents the initial angles that  satisfy the condition that
  $\dot \theta(\tau)$ equals $ \dot \theta_0 $, the initial angular
  velocity of $\theta$. Fortunately, the contours $\delta \alpha =
  0$ meets  $\dot \theta = \dot \theta_0$ lines. However, at that
  initial conditions, the final condition that makes the setup
  perfect for the next turn is not satisfied. The solid blue line
  in  Fig. \ref{FigContourA}, represents the initial angles which gives the condition
  $\dot \alpha (\tau) = 0$. Unfortunately, there are no
  initial angles that satisfy $\delta \alpha = 0$, $\dot \theta =
  \dot \theta_0$, and $\dot \alpha(\tau) = 0$ simultaneously. This
  means that without any extra force, there is no successive ski turn
  in our model. However, in actual ski movement, the skier uses two ski plates,
   so it is very easy to control the angle $\alpha$, and the poles make it
   even easier to control the angular movement of  $\alpha$.

   From now on, we carefully find the critical condition $\dot \theta (\tau) = \dot \theta_0 $ in order to
   make  successive  ski turns. For the $\alpha$ angle, we consider
   two cases, such as $\alpha (\tau) =  0$ and $ \alpha (\tau) =
   -\alpha_i $. Although,  $ \alpha (\tau) =    -\alpha_i $ is the correct one,
   usually at that condition,  the amplitude of the $\dot \alpha (\tau)$ is very
   big and the generalized force $Q_z$ has very high negative values. From the skier's view-point,
   the skier needs large body movement, including  pole action, to follow a circular  trajectory.
   We think that the condition for $\alpha(\tau) =0$ needs milder
   movement of the skier to keep the circular trajectory. Therefore,
   in our calculation, we consider  two cases $\delta \alpha (\tau) = - \alpha_i $
   and $\alpha (\tau) = 0$   with  $\dot \theta (\tau) = \dot \theta_0 $ in order to make successive ski turns.
 We also plotted the contours line that represent the initial
 angles  that gives the generalized force at $ \theta = 90^{\circ} $
 that becomes $Q_z = 0$ in  Fig. \ref{FigContourA}.

\begin{figure}[htbp]
\centering
\includegraphics[width=10cm]{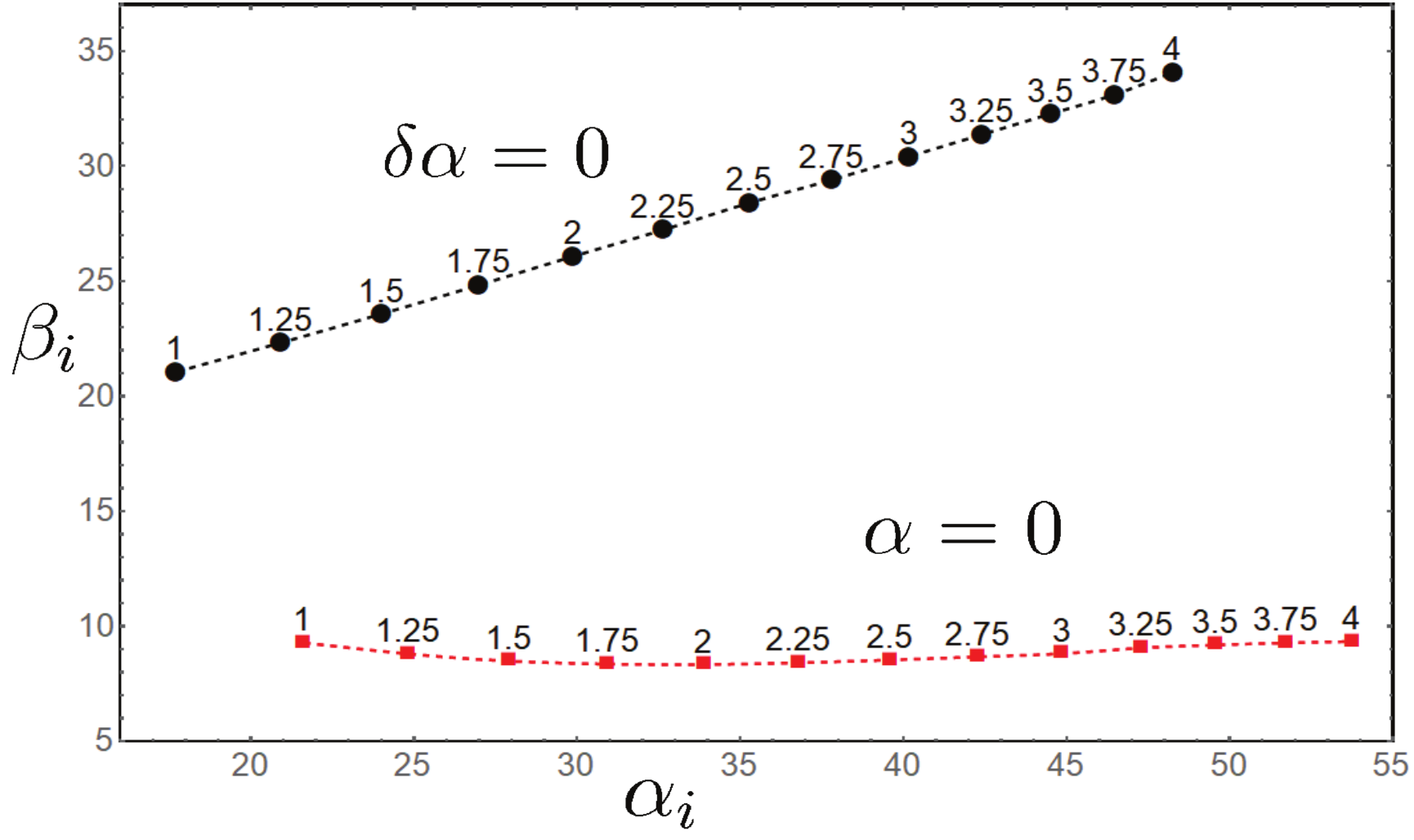}
\caption{Each point represent the initial angles $\alpha$ and
$\beta$ that give the final condition such as $\delta \alpha = 0$
and $\alpha=0$. The numbers from $1$ to $4$ on each point indicate
the radius ($ r_0$)  of the circular path of the ski trajectory. The
initial angular velocity $\dot \theta_0$ is fixed for all $r_0$
values. } \label{FigRadius}
\end{figure}

The condition for successive ski turns can be changed as the radius
of the circular trajectory. In Fig.  \ref{FigRadius}, we plotted the
initial angles $\alpha_i $ and $\beta_i$ that give the condition
$\delta \alpha (\tau) = 0$, or $\alpha (\tau) = 0$ at $\theta = 90
^{\circ}$. At this time each point satisfies the condition $\dot
\theta (\tau) = \dot \theta_0 $, where $\dot \theta_0 = \frac{120
\pi}{180}$. We fixed the slope angle $\phi = 20^{\circ}$, and
changed the radius($ r_0$) of the circular trajectory from $1 {\rm
m}$ to $4 {\rm m} $. As the radius increase, both the initial angles
$\alpha_i $ and $\beta_i$ increase. From the skier's view, it gives
counter intuitive result at first glance. Since the skier moves
generally more actively as the radius of the turn becomes smaller,
in other words, the skier leans her/his body more forward and more
inside of the turn as the radius of the turn becomes smaller.
Actually, in our calculation, we fixed not the linear velocity, but
the initial angular velocity $\dot \theta_0$. In other words, as the
radius of the turn becomes larger, the linear velocity also becomes
large. Therefore, the results in Fig. \ref{FigRadius} shows that as
the skier moves fast, the skier leans further forward (bigger
$\beta_i$)and more inside  of the turn (bigger $\alpha_i$).  The
interesting thing is that for the condition $\alpha(\tau) = 0$, the
initial angle $\beta_i$ is not changed very much, the skier only has
to control the initial angle $\alpha_i$ as the speed of the ski
becomes large. Considering the $\dot \alpha (\tau)$ and the
generalized force $Q_z$, the skiers initial condition had better
stay between the two lines in Fig. \ref{FigRadius}.

\begin{figure}[htbp]
\centering
\includegraphics[width=10cm]{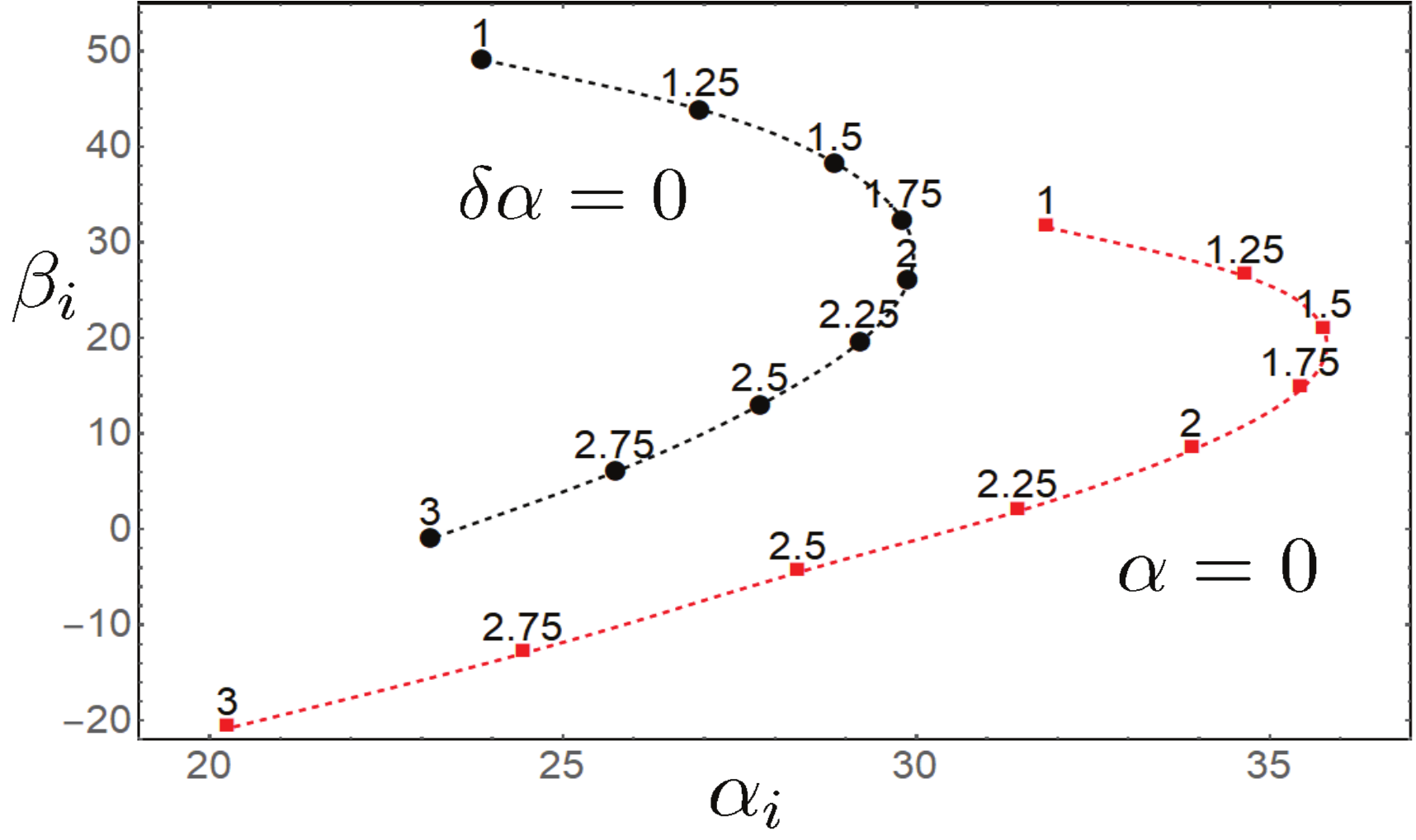}
\caption{Each point represent the initial angles $\alpha$ and
$\beta$ that give the final condition such as $\delta \alpha = 0$
and $\alpha=0$. The numbers from $1$ to $4$ on each point indicate
the radius ($ r_0$)  of the circular path of the ski trajectory. The
initial linear velocity of $ r_0 \dot \theta_0 $ is fixed for all
$r_0$ values.} \label{FigRadiusOR}
\end{figure}

Instead of keeping the angular velocity constant, we keep the linear
velocity constant, and we find the initial condition to make
successive turn as a function of the radius of the circular
trajectory.
 In Fig.  \ref{FigRadiusOR}, we plotted the
initial angles $\alpha_i $ and $\beta_i$ that give the condition
$\delta \alpha (\tau) = 0$, or $\alpha (\tau) = 0$ at $\theta = 90
^{\circ}$. At this time, each point satisfies the condition $\dot
\theta (\tau) = \dot \theta_0 $, where $\dot \theta_0 $ is  not a
constant for various radius. We fix the linear velocity $v_0 = 2
\frac{120 \pi}{180} $ and the angular velocity is a function of the
radius $ r_0$,  $\dot \theta_0 = \frac{v_0}{r_0}$. The slope angle
is fixed as  $\phi = 20^{\circ}$ and  the radius($ r_0$) of the
circular trajectory  is changed from $1 {\rm m}$ to $4 {\rm m} $.

As the radius of the circular trajectory decreases, the initial
angle $\beta_i$ increases. From the skier's view-point, it is very
natural that the skier should lean forward to make a short radius
turn with the same initial linear velocity.
 The interesting thing happens with the initial angle $\alpha_i$ when we get an condition
  such as $\delta \alpha (\tau) =0$,  as the radius $r_0$ increases from $1 {\rm m}$,
   the initial angle $\alpha_i$ increases at first,
  but  when the radius $r_0$ is greater than $2 {\rm m}$ the initial angle $\alpha_i$ decreases.
If we consider the condition
  such as $ \alpha (\tau) =0$, there is a similar trend in the initial
  conditions.

  From the skier's view-point, the skier should lean forward and
  inward of the circle as the linear velocity increases. This is the
  case  when the radius changes from $3 {\rm m}$ to around  $2 {\rm m}$ in
   Fig. \ref{FigRadiusOR}. However, the radius is comparable to the
   length $l_0$ (between the ski and point mass), and the motion of the point mass
   is not so simple. The numerical result shows that the $\alpha_i$ should decrease
   as the angular velocity gets higher in some region.

\begin{figure}[htbp]
\centering
\includegraphics[width=10cm]{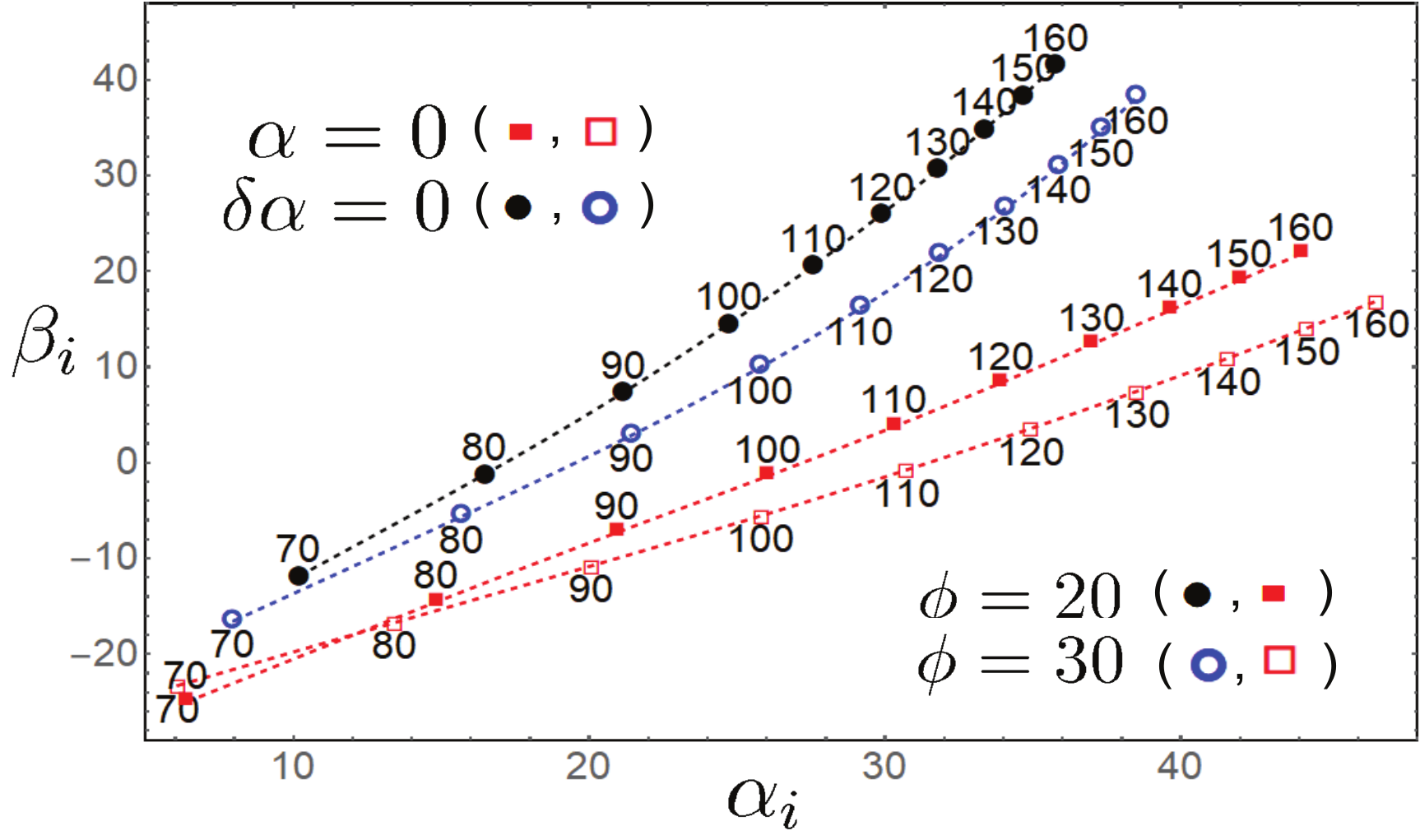}
\caption{ The numbers from $70$ to $160$ on each point indicate the
initial angular velocity of $\theta$ from $\dot \theta_0 = \frac{70
\pi}{180}$ to $\dot \theta_0 = \frac{160 \pi}{180}$. Filled (open)
circles represent the case  the  initial angles $\alpha$ and $\beta$
that give the final condition such as $\delta \alpha = 0$ for the
slope angle $\phi = 20^{\circ} $ ($\phi = 30^{\circ} $). Filled
(open) squares represent the case where the  initial angles $\alpha$
and $\beta$ give the final condition $\alpha = 0$ for  the slope
angle $\phi = 20^{\circ} $ ($\phi = 30^{\circ} $).
 } \label{FigPhi2030}
\end{figure}

Keeping the radius of the circular trajectory($r_0 = 2{\rm m}$), we
change the angular velocity $\dot \theta_0$ from $\frac{70 \pi}
{180}$ to $\frac{160 \pi}{180}$ in Fig. \ref {FigPhi2030}. Filled
black circles represent the case where the  initial angles $\alpha_i
$ and $\beta_i$  give the final condition such as $\delta \alpha
(\tau) = 0$ for the slope angle $\phi = 20^{\circ} $  The condition
$ \alpha (\tau) = 0$ is also shown in Fig.  \ref {FigPhi2030} as the
filled red squares. As we expect,  as the angular velocity $\dot
\theta_0$ increases, the skier should lean more forward and towards
the center  of the circle. We also find the initial angles for the
slope angle  $\phi = 30^{\circ} $. The filled circle represent the
initial angles for the slope  $\phi = 20^{\circ} $ and the empty
circle does so for the slope  $\phi = 30^{\circ} $, which satisfy
the condition $\delta \alpha (\tau) = 0$. Considering the slope
angle, one might think that the skier should make a larger angle
toward the center of the turn, and lean forward with an large angle.
Actually, the graph in  Fig. \ref {FigPhi2030} shows what we expect.
What we missed at first is the fact that the angles are defined from
the $z$-axis, and the $z$-axis is normal to the slope. Therefore, as
the slope increase, the $z$-axis already moves far from the $-
\vec{g} $ direction by the increasing angle. From the skier's
view-point, he/she leans more severely towards the fall line
direction. In other words, the initial angle for the slope   $\phi =
20^{\circ} $is generally greater than the initial angle for the
slope $\phi = 30^{\circ} $. However, this does not mean that the
skier leans more severely to the fall line as the slope angle
increases.

In this section, we solved the differential equations (Eq.
\ref{EqLambdaTheta} - \ref{EqLambdaBeta}) numerically, with setting
  angle $\beta$  as a constant value.  The initial condition of
two angles $\alpha$ and $\beta$ is studied for both making  a
complete circular turn without falling, and making a successive
turn. In general, there is no initial condition that the final
condition at $\theta = 90^{\circ}$ is another new initial condition
for the next turn. Since the skier can adjust his/her motion by
his/her two legs and poles,  we checked the angular velocity of
$\theta$ at first, in order to make the same initial condition for
the next turn in this section.

\section {Effect of the skier's active Movement. } \label{secAvtive}

On the slope, a skier usually actively move his/her body in order to
maintain stable position and control the ski path. In section
\ref{secExample}, our model moves passively during the ski turn, in
other words, once the initial angles $\alpha_i$ and $\beta_i$ are
chosen, the point mass $m$ on the bar  moves down hill with the same
angle $\beta_i$ along the circular ski trajectory without any other
active movement. The equations of motion determine the motion of the
point mass $m$ and give the final angle $\alpha_f$ and the angular
velocity $\dot \theta(\tau)$.

 In this section, we add active movement on the angle $\beta$, which relates the skier's motion
 along the ski plates. We set
\begin{eqnarray}
\beta(t) = b_0 + b_1 \cos(\omega t+ \psi),  \label{EqBeta}
\end{eqnarray}
where $\omega = \frac{2 \pi}{\tau}$, and $\tau$ is the time
$\theta(\tau) = 90^{\circ}$ for each turn. The $\omega$ is roughly
twice the frequency of the ski turn. In other words, the skier moves
back and forth with amplitude $b_1$ and makes complete motion during
the time from $\theta = - 90 ^{\circ}$ to $\theta = 90^{\circ}$.
With $b_1 = 0$, the skier motion is the same as the result discussed
in section \ref{secExample}.

\begin{figure}[htbp]
\centering
\includegraphics[width=10cm]{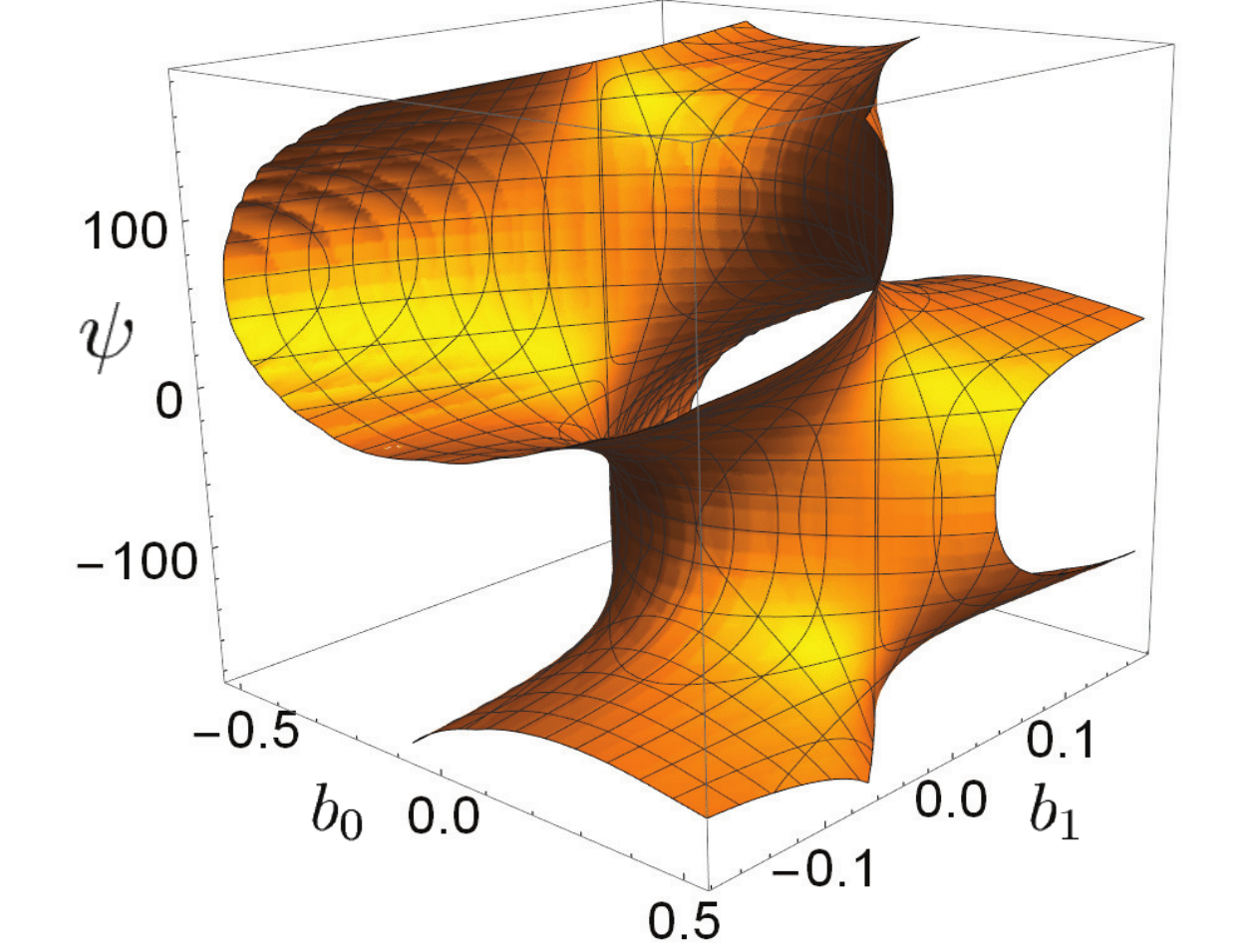}
\caption{Contour surface indicating the initial values of $b_0$,
$b_1$ and $\psi$ that give the angle $\alpha = 0$ at $\theta =
90^{\circ}$. The initial angle of the $\alpha$ is $\alpha_0 =
32.5^{\circ}$, with the slope angle $\phi = 30^{\circ}$. }
\label{FigCon32}
\end{figure}

\begin{figure}[htbp]
\centering
\includegraphics[width=10cm]{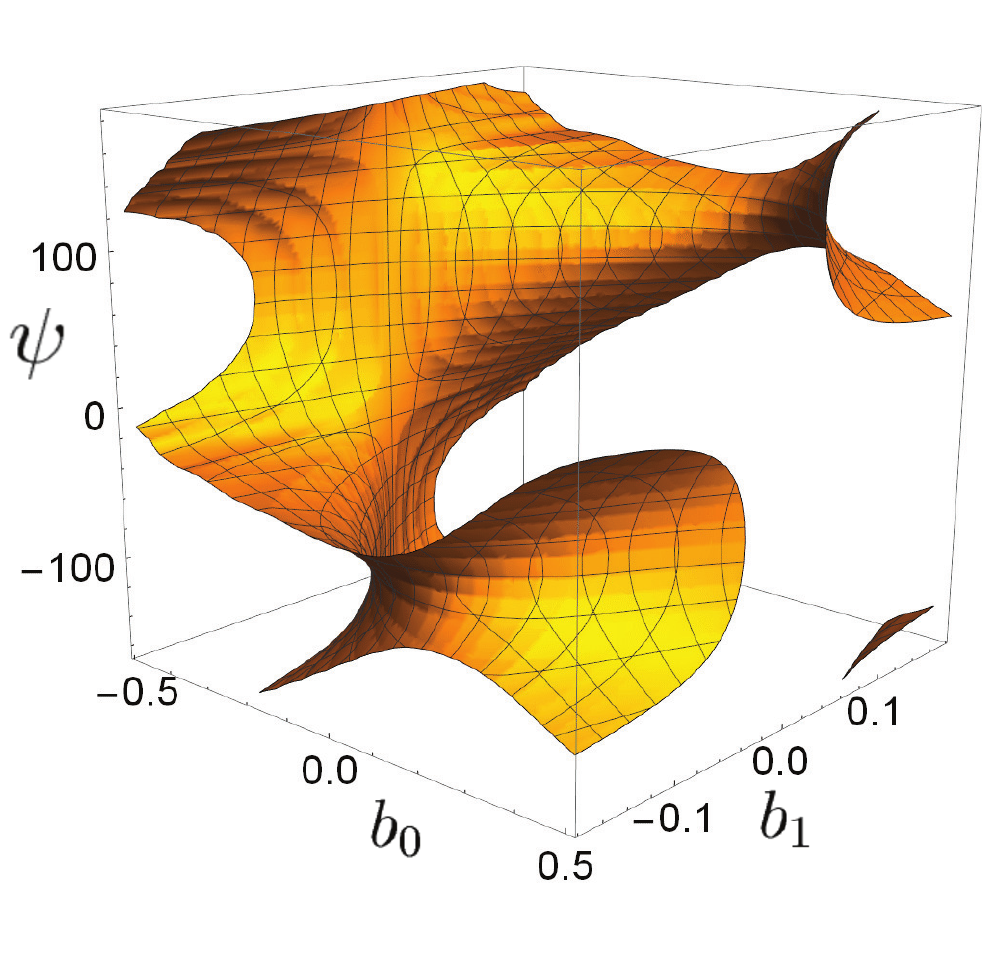}
\caption{ Contour surface indicates the initial values of $b_0$,
$b_1$ and $\psi$ that give the angle $\alpha = 0$ at $\theta =
90^{\circ}$. The initial angle of the $\alpha$ is $\alpha_0 =
29^{\circ}$, with the slope angle $\phi = 30^{\circ}$. }
\label{FigCon29}
\end{figure}

  In Fig. \ref {FigCon32},  the contour
surfaces represent the ($b_0, b_1, \psi$) set that gives
$\alpha(\tau) = 0$ with $\dot \theta(\tau) = \dot \theta_0 $. With
the same slope angle ($\phi = 30^{\circ}$, we set the initial angle
of $\alpha$ from $32.5^{\circ}$ to $29^{\circ}$, then the contour
surface changes as in Fig. \ref {FigCon29}. These figures show that
the active movement on the angle $\beta$ may give various motion of
the ski.


\begin{figure}[htbp]
\centering
\includegraphics[width=10cm]{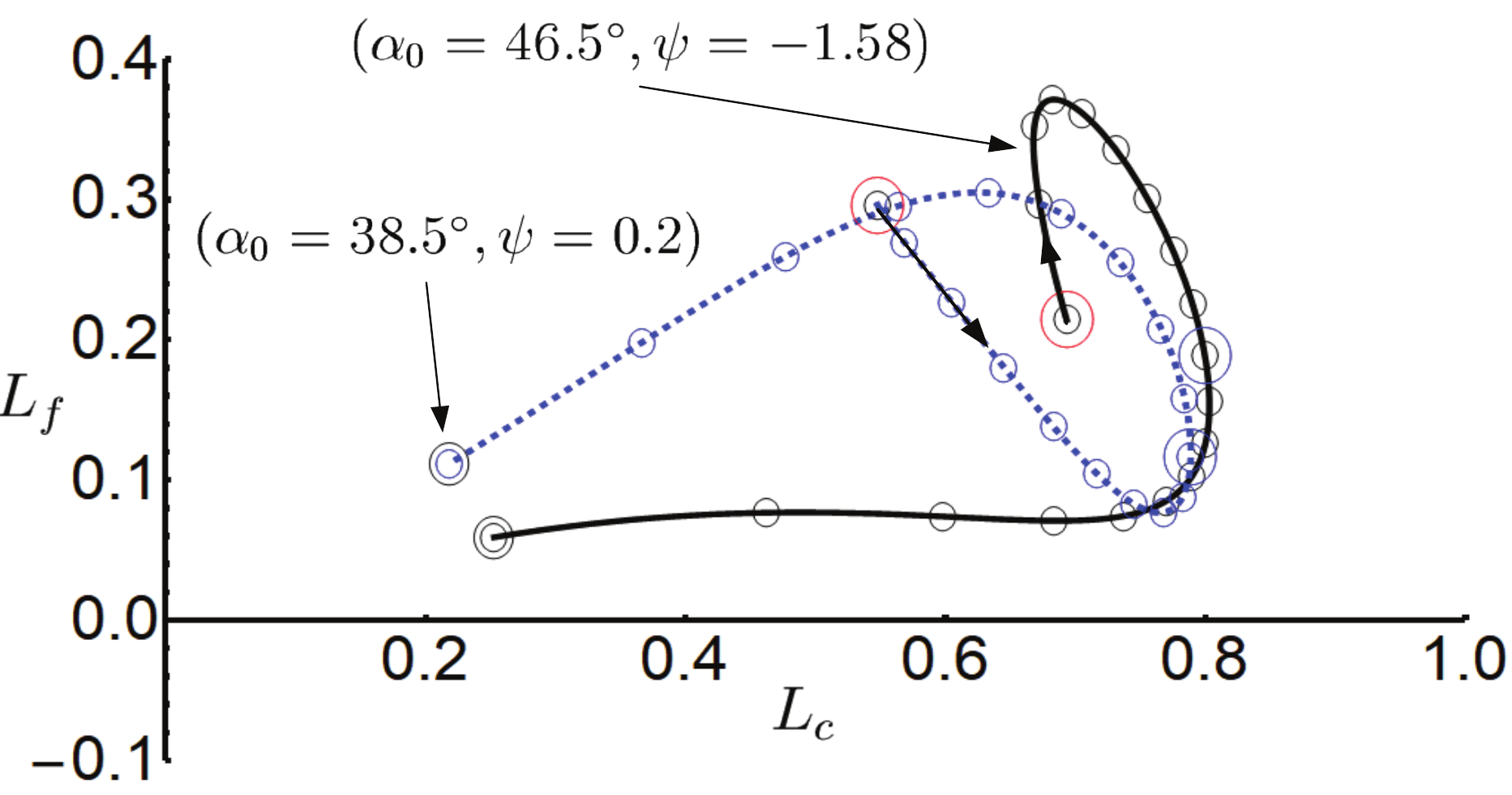}
\caption{Traces of the point mass $m$ as the $\theta$ increase by
$10^{\circ}$. $L_c$ is the projected length to the ground toward the
center of the circular orbit. $L_f$ is the projected length to the
ground toward the forward direction of ski movement. The dashed line
moves counterclockwise and the solid line moves clockwise.
 Red circles represents the starting point ($\theta = -90^{\circ} $). .
$( b_0 = 0.3 , b_1 = 0.2 )$} \label{FigHeadMoveCCW}
\end{figure}

Figures \ref{FigHeadMoveCCW} shows the position of the point mass
$m$ on the rod ($l_0 = 1 {\rm m}$), which connects the point mass
$m$ and the ski plate. $L_c$ is the projected length to the ground
toward the center of the circular ski trajectory. $L_f$ is the
projected length to the ground towards the forward direction of the
ski movement. For the same slope, with the same constant $b_0 = 0.3,
b_1 =0.2 $, we plot the trace of the point mass by a solid line and
dashed line for the set $(\alpha_0 = 38.5 ^{\circ}, \psi = 0.2 ) $
and $(\alpha_0 = 46.5 ^{\circ}, \psi = -1.58 )$, respectively. The
red circle indicates the point mass at the initial position and the
blue circles indicates the point mass when the ski is parallel to
the fall line, i.e. $\theta = 0 ^{\circ}$.

 The phase factor $\psi$
in  Eq. \ref{EqBeta} makes the point mass turn clockwise or
counterclockwise. Each circle in the solid line indicate the trace
of the point mass per $10^{\circ}$ in the angle $\theta$. Following
the solid line, the skier leans forward till he/she moves till
$\theta = -60^{\circ}$, then he/she leans relatively backward until
he/she reaches $\theta=90^{\circ}$. In this set up, the final
angular velocity $\dot \theta (\tau) = 3.14$ and it takes $0.80 s$.
On the other hand, if we follow the dashed line, the point mass
moves counter clockwise. With the initial angle $\alpha_i = 38.5$
and the phase $\psi = 0.2 $, the skier leans relatively backward
from the initial condition till he/she reaches $\theta =
-20^{\circ}$, almost until the ski is parallel to the fall line,
then he/she moves forward till he/she reaches $\theta = 50^{\circ}$,
after that he/she slowly leans backward to finish his/her turn. In
this case, the final angular velocity $\dot \theta (\tau) = 3.78$
and it takes $0.89 {\rm s}$. Comparing this case with the clock wise
movement, it makes the final angular velocity much greater and it
takes longer time. Generally, the skier lean forward at the first
part of the turn and eventually he/she  moves back at the end of the
turn.

%

\begin{figure}[htbp]
\centering
\includegraphics[width=10cm]{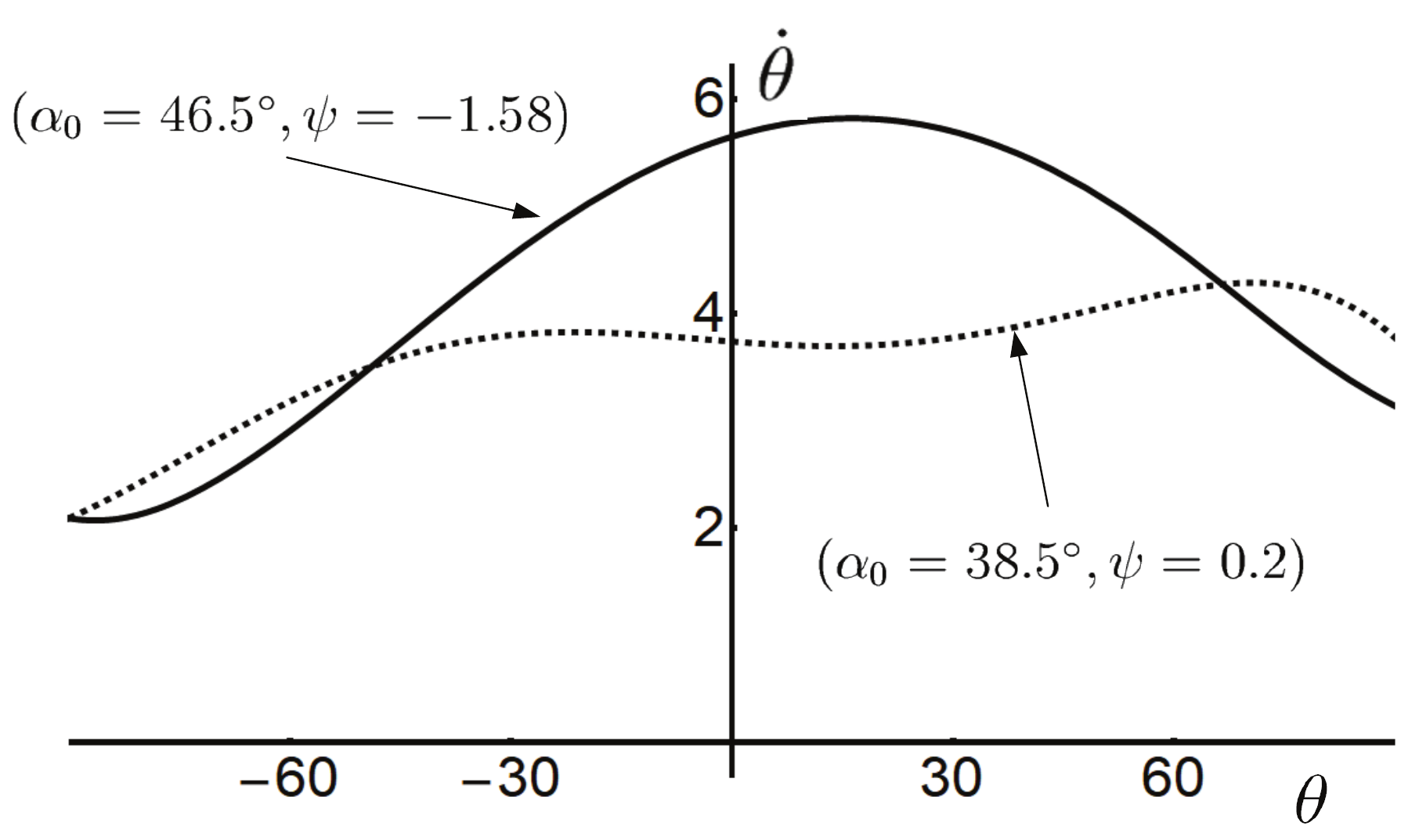}
\caption{The solid lines represent the angular velocity $\dot
\theta$ as a function of $\theta$  for the condition  $(\alpha_i =
46.5 ^{\circ}, \psi = -1.58) $.
   The dashed lines represent the angular velocity $\dot
\theta$ as a function of $\theta$  for the condition $(\alpha_i =
38.5 ^{\circ}, \psi = 0.2) $.
 }
\label{FigThetaP}
\end{figure}

In Fig. \ref{FigThetaP}, we plot the angular velocity $\dot \theta$
as a function of $\theta$  for the case  $(\alpha_i = 46.5 ^{\circ},
\psi = -1.58) $, and $(\alpha_i = 38.5 ^{\circ}, \psi = 0.2) $. For
the solid line, the point mass $m$ moves clockwise, the angular
velocity $\dot \theta$ increases just after the fall line, and it
decreases as the point mass reaches the end of the turn ($\theta =
90^{\circ}$). On the other hand, the dashed line, the point mass $m$
moves counter clockwise, the angular velocity does not increase as
much as the clockwise case at the fall line, but the angular
velocity does not decrease and it stays as a higher value. This
means that, at the end of the turn, the skier still has higher
velocity, and it is not easy to control his/her body to start a new
turn. The worst thing is that the skier takes more time to complete
one turn even though his/her final velocity is higher.

\begin{figure}[htbp]
\centering
\includegraphics[width=10cm]{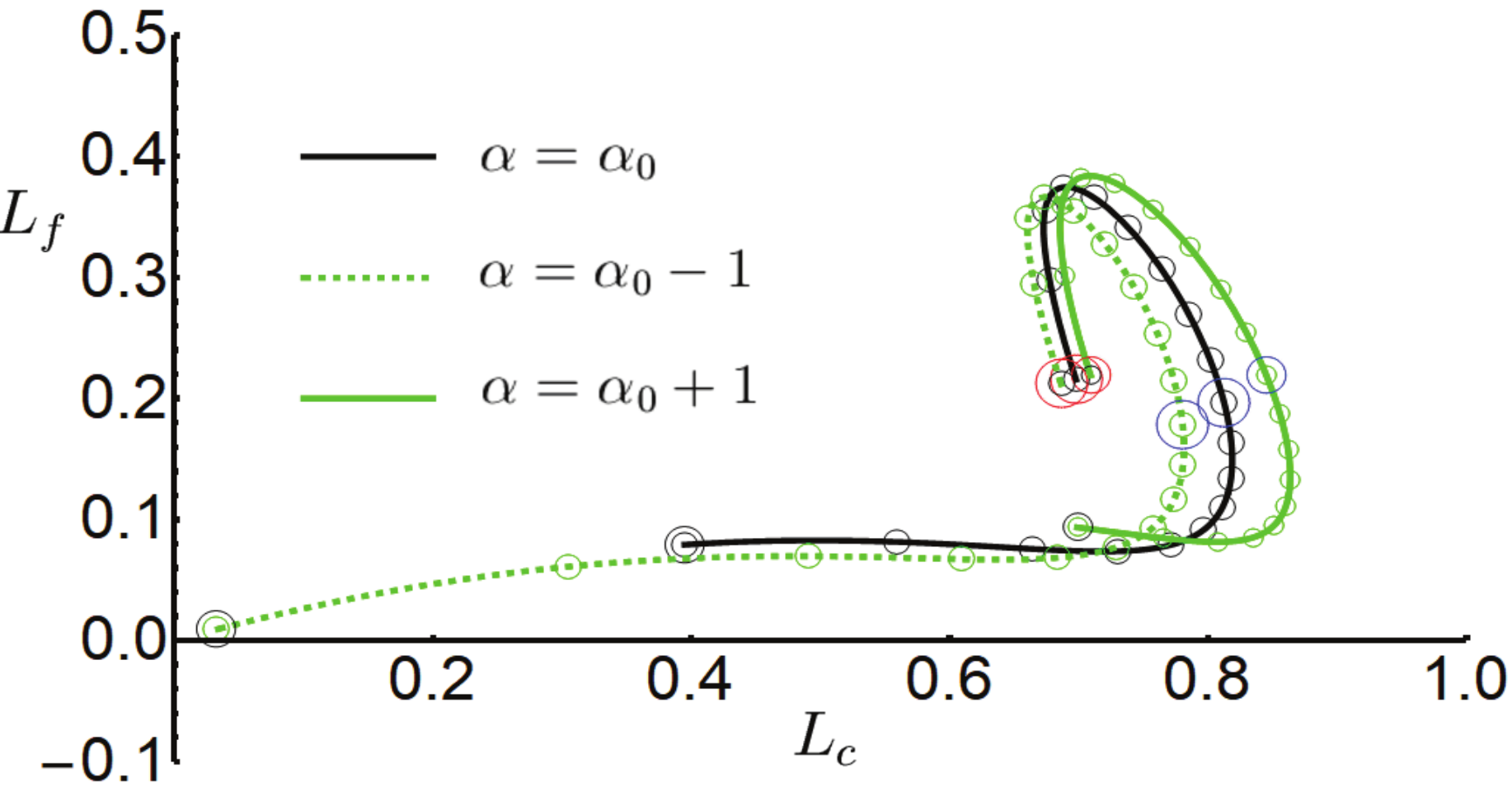}
\caption{Traces of the point mass $m$ as the $\theta$ increase by
$10^{\circ}$. $L_c$ is the projected length to the ground toward the
center of the circular orbit. $L_f$ is the projected length to the
ground toward the forward direction of  ski movement.   Red circles
represents the starting point ($\theta = -90^{\circ} $) and the blue
circle and the black circle represents the mid point($\theta =
0$)and the final points($\theta = 90^{\circ}$), respectively.
$\alpha_0 =47^{\circ}$. $( b_0 = 0.3 , b_1 = 0.2, \psi = -
\frac{\pi}{2}$
 } \label{FReleaseTime}
\end{figure}

\begin{figure}[htbp]
\centering
\includegraphics[width=10cm]{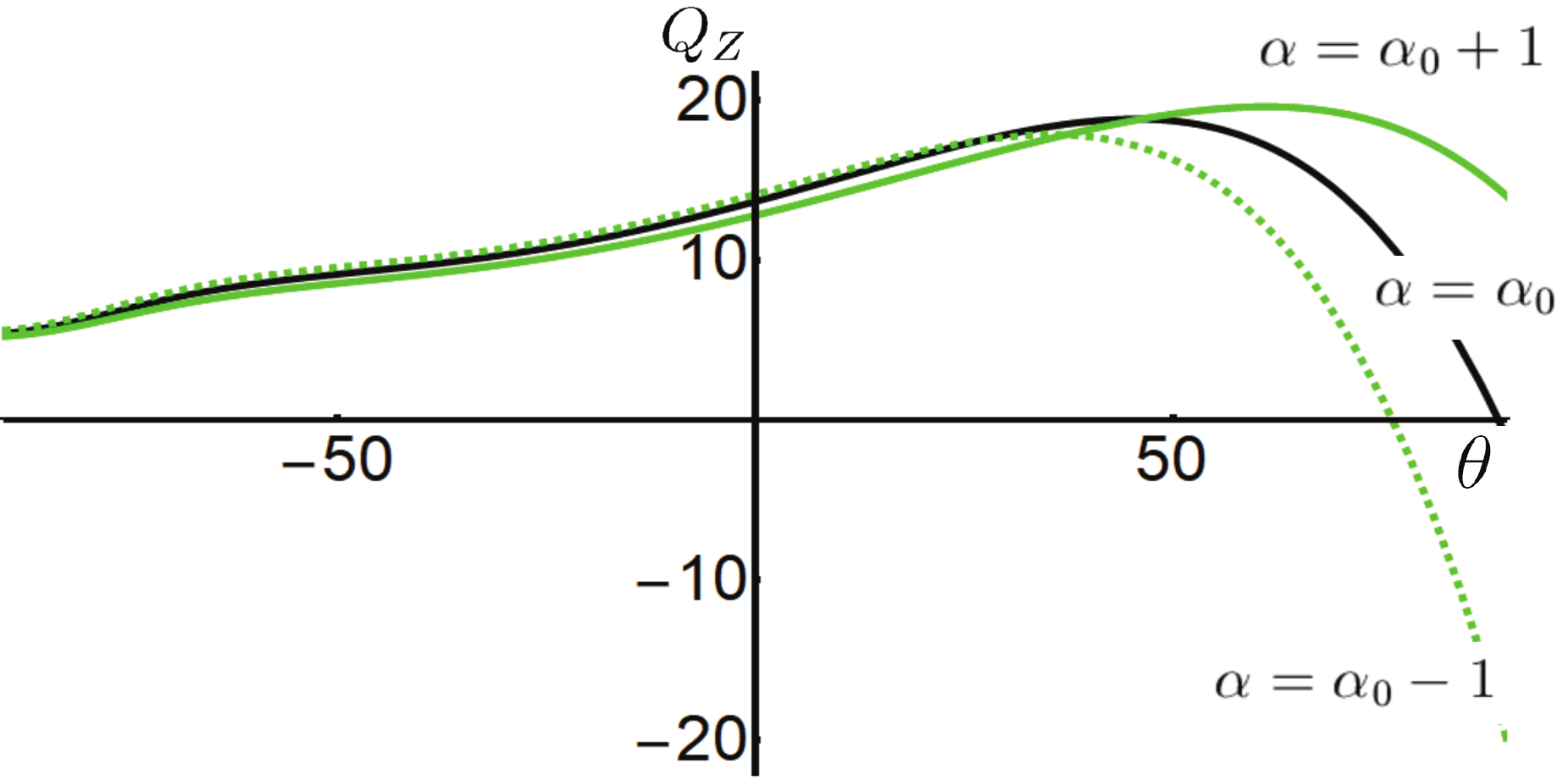}
\caption{The solid black line represents the generalized force $Q_z$
with the initial angle $\alpha_0 =47^{\circ}$. The solid(dashed)
green line represents the generalized force $Q_z$ with the initial
angle $\alpha_0 + 1^{\circ}$ ($\alpha_0 + 1^{\circ}$).
 }
\label{FigQZ}
\end{figure}

In Fig. \ref{FReleaseTime}, we plot the position of the point mass
$m$ on the rod ($l_0 = 1 {\rm m}$) that connects the point mass $m$
and the ski plate.  The solid black line, the dashed green line, and
the solid green line represent the traces of the point mass for
$\alpha$ = ($ 47^{\circ},46^{\circ}, 48^{\circ}$), respectively. The
red circle indicates the initial position and the blue circles
indicate the point mass when the ski is parallel to the fall line,
i.e. $\theta = 0 ^{\circ}$. With the small change of the initial
angle $\alpha$  by $1^{\circ}$, the final angle $\alpha_f$ changes a
lot. If we increase the initial angle, then the final position of
the point mass stays close to the center of the circular trajectory.
In contrast, if we decrease the initial angle $\alpha_i$, the point
mass return back to the upright position i.e. $\alpha_f \sim 0$. The
changes of generalized force $Q_z$ in Fig. \ref{FigQZ} show the
differences caused by the change of initial angle $\alpha_i$. With
the initial angle $\alpha_i = \alpha_0$, the generalized force $Q_z$
becomes $0$ at $\theta = 90^{\circ}$. The generalized force stay
positive at  $\theta = 90^{\circ}$  with the initial angle $\alpha_i
= \alpha_0 +1^{\circ}$. This condition makes the point mass $m$ keep
follow the
 circular trajectory upward against the slope,then the skier has s
 hard time to initiate another downward turn.
  With $\alpha_i = \alpha_0 - 1^{\circ}$,
  the generalized force
 becomes zero before the ski arrives $\theta=90^{\circ}$, and it
 becomes negative at  $\theta=90^{\circ}$. From the skier's view,
 he/she
 obtains a repulsive force from the ground, and this force enables
 him/her to  move upward. He/shw may use this force in order to adjust his/her body
 to initiate a new ski turn from that position.


%

\begin{figure}[htbp]
\centering
\includegraphics[width=10cm]{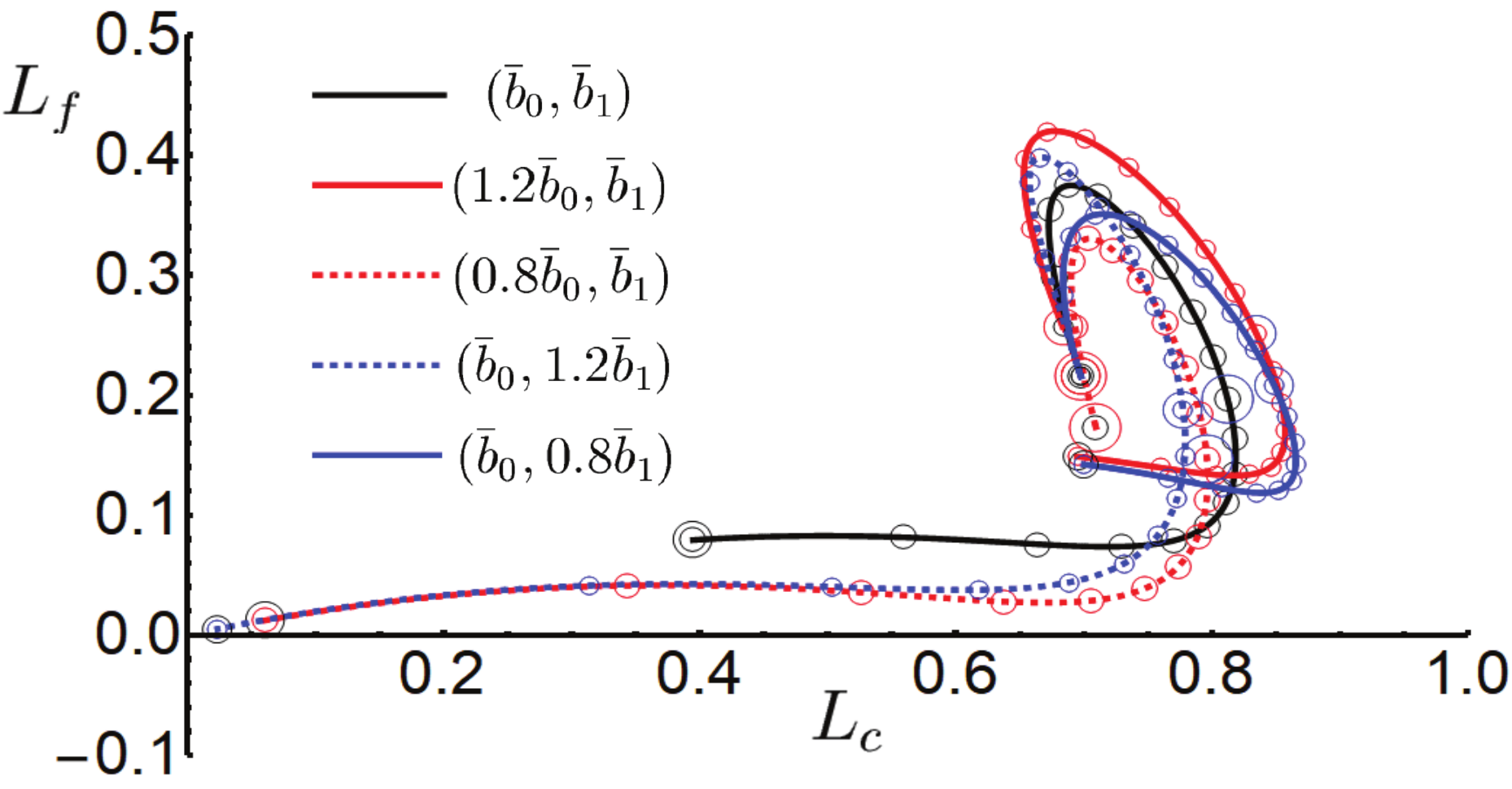}
\caption{Traces of the point mass $m$ as the $\theta$ increase by
$10^{\circ}$. $L_c$ is the projected length to the ground toward the
center of the circular orbit. $L_f$ is the projected length to the
ground toward the forward direction of ski movement. The solid black
line is for the case $b_0 = \bar b_0$ and $b_1 = \bar b_1$. The
initial angle $\alpha$ for all the cases is the same as $\alpha_0$.
 } \label{FigPbpme}
\end{figure}

\begin{figure}[htbp]
\centering
\includegraphics[width=10cm]{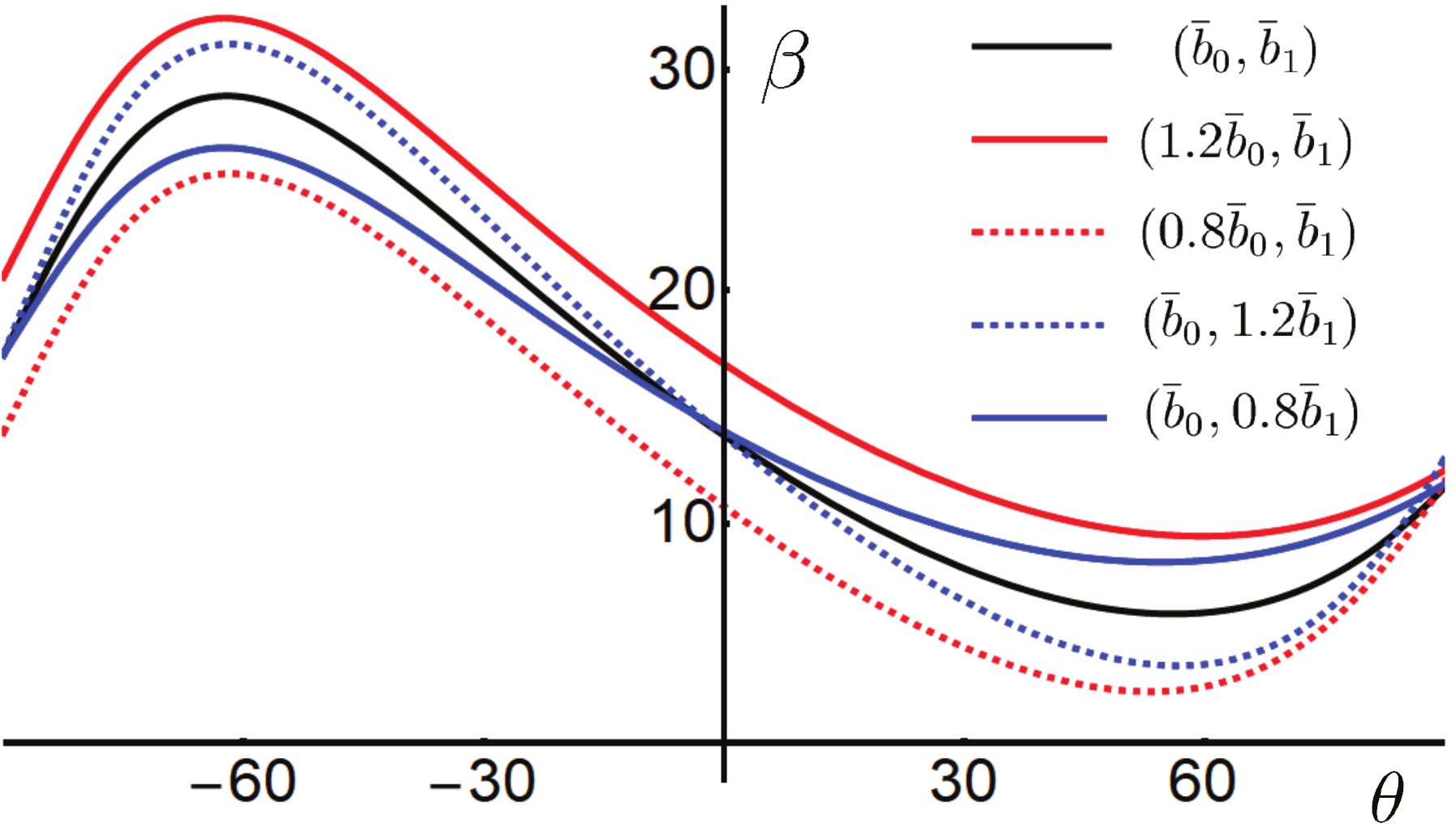}
\caption{Time dependence of the angle $\beta$.  The solid black line
is for the case $b_0 = \bar b_0$ and $b_1 = \bar b_1$, where $\bar
b_0 = 0.3, \bar b_1 = 0.2$.
 } \label{FigBeta}
\end{figure}

In Fig. \ref{FigPbpme}, we plot the position of the point mass $m$
as functions of $b_0$ and $b_1$ in Eq. \ref{EqBeta}. The solid black
line can be a reference trace as in Fig. \ref{FReleaseTime} with
$(\alpha_0 =47^{\circ}, b_0 = \bar b_0, b_1 = \bar b_1 ,  \psi = -
\frac{\pi}{2}) $,where $\bar b_0 = 0.3, \bar b_1 = 0.2$. The actual
movement of the angle $\beta$ is in Fig. \ref{FigBeta}.
 The solid red line represents the trace with $b_0 = 1.2 \bar b_0$, and
the dashed red line represents the trace with $b_0 = 0.8 \bar b_0$.
 The two traces show that, if we decrease the constant angle $b_0$ by a
little, then the trace of the point mass $m$ becomes easier to adapt
to the next turn. From the skier's view-point, the  body can be
positioned upright to the ski plate.

The dashed  blue line represents the trace with $b_1 = 1.2 \bar
b_1$, and the solid blue line represents the trace with $b_1 = 0.8
\bar b_1$. The value $b_1$ is related to the modulation amplitude of
the $\beta$ movement.  If the modulation amplitude is smaller than
$b_1$, the solid blue line shows that the trace of the point mass
$m$ is much closed to the solid red line at $\theta = 90^{\circ}$

The position of the point mass $m$ around the final turn $\theta =
90^{\circ}$ in Fig. \ref{FigPbpme} shows two groups. The first one
is the dashed blue and dashed red lines, while the other is the
solid red and solid blue lines. These two groups can be explained in
the Fig. \ref{FigBeta} only if we consider the angle $\beta$ after
$\theta = 60^{\circ}$. In other words, the important $\beta$
movement in the ski turn happens  after the skis cross the fall
line, in other words $\theta$ is greater than $30^{\circ}$.

In this section, the angle $\beta$ is not a constant in solving the
differential equations (Eq. \ref{EqLambdaTheta} -
\ref{EqLambdaBeta}) numerically. We set the angle $\beta$ as a
time-dependent function, which can duplicate the skier's motion
along the ski plate back and forth. We analyzed the motion of the
point mass $m$ which corresponds to the center of the skier's mass
as the angle $\beta$ changes as a function of time.

\section{Conclusion and Discussion.} \label{conclusion}

    We studied the skier's motion based on a simple model, in which the
point mass $m$ on a  single rod is connected to a single ski plate.
We at first excluded the skier's active motion  and find the
particular solution for the point mass $m$ to make stable ski turns.
Our model neglected the intrinsic motion of the ski plate; we simply
assumed that the single massless ski plate moves along a circular
trajectory on an inclined plane slope. For a certain initial angle
$\beta_i$,  we can find a time-dependent angle $\alpha(t)$, which
gives  the point mass $m$ a complete half-turn without falling.

The generalized force ($Q_z$) supporting the point mass $m$ from the
ski plate was calculated. A skier can control  $Q_z$  by up and down
movement. However, in our model  $Q_z$ is calculated without any
skier's active movement. $Q_z$  can be explained by a rebound force
from the ground. This rebound force is not related to
 any geometrical structure of the ski plate.

Although we can't find a complete solution for  successive turns, a
final condition at $\theta = 90^{\circ}$ gives a new solution for
the initial angles making a next half turn without falling. In an
actual ski turn, the skier uses two legs and poles to adjust the
movement of the angle $\alpha$;  we tried to find a solution with
some tolerance for the angle $\alpha$. As the degree of the slope
increases, the initial angles $\alpha_i$ and $\beta_i$ should
increase for a stable ski turn. We studied the conditions for the
stable ski turn as functions of the linear velocity and radius of
the turn. From the skier's view-point, the solutions for the stable
ski turn  do not require any extra movement to complete a stable
circular turn. Then the solution may give the skier the
   most comfortable skiing method without any active movement to
  control the ski.

 In an actual ski turn, the skier may add extra movement along the ski
 plate ($\vec t $-direction). Adding an active movement to the direction
 of the ski plate,   the conditions for the stable ski turn were analyzed.
 With active fine tuning of the angle $\beta$, the motion of the
 point mass $m$ was studied in detail. The final angle $\alpha_f$
 depends on the angle  $\beta$ after passing through the fall line ($\theta = 0 $).

 In our study, we used simple model for the skier's movement,
  so the results are not directly applicable to actual skiing
  process.  However, our result gives some insight into  the skier who wants to
  develop his/her technique.

\end{document}